\documentclass[preprint,aps,nofootinbib,11pt]{revtex4-1}
\usepackage{geometry}                
\usepackage{amsmath,amssymb,amsfonts,dcolumn,color,graphicx,graphics,latexsym,placeins,epsfig,tikz}
\usetikzlibrary{arrows,shapes}
\usepackage{subfigure,rotating,bm,mathrsfs}
\usepackage[title]{appendix}

\newcommand{\be}{\begin{equation}}
\newcommand{\ee}{\end{equation}}
\newcommand{\ba}{\begin{eqnarray}}
\newcommand{\ea}{\end{eqnarray}}

\usepackage[pagebackref=false, colorlinks=true]{hyperref}
\definecolor{redish}{rgb}{0.7,0.2,0.0}  
\definecolor{bluish}{rgb}{0.2,0.5,0.8}

\hypersetup{linkcolor=redish,          
                  citecolor=blue,        
                  filecolor=magenta,      
                  urlcolor=blue}          

\begin{document}

\author{Rajibul Shaikh}
\email{lrajibulsk@gmail.com}
\affiliation{Institute of Convergence Fundamental Studies, School of Natural Sciences, College of Liberal Arts, Seoul National University of Science and Technology, Seoul 01811, Korea}

\title{\Large Testing black hole mimickers with the Event Horizon Telescope image of Sagittarius A$^*$}

\begin{abstract}
The Event Horizon Telescope (EHT) has recently observed the image and shadow of the supermassive compact object Sagittarius A$^*$ (Sgr A$^*$). According to the EHT collaboration, the observed image is consistent with the expected appearance of a Kerr black hole. However, it is well-known that some non-Kerr objects may mimic many of the properties of the Kerr black hole, and hence, their shadows might be consistent with the observed shadow of Sgr A$^*$. In this work, we consider two black hole mimickers and study their shadows. The first mimicker is a rotating generalisation of the recently proposed static, spherically symmetric black-bounce spacetime by Simpson and Visser where the central Schwarzschild singularity is replaced by a minimal surface. The second one is the $\gamma$-metric which is a static, axially symmetric singular solution of the vacuum Einstein's equations without an event horizon. We put constraint on the parameters of these black hole mimickers by comparing their shadows with the observed shadow of Sgr A$^*$.

\end{abstract}

\maketitle
\section{Introduction}
The recent observations of the images and shadows of the supermassive compact object Sagittarius A$^*$ (Sgr A$^*$) at the heart of Our Galaxy \citep{SgrA_EHT1,SgrA_EHT2} as well as of the compact object M87$^*$ at the heart of the nearby galaxy M87 \citep{M87_EHT1,M87_EHT2,M87_EHT3} by the Event Horizon Telescope (EHT) collaboration have opened up a new window in observational astronomy to probe and test gravity and fundamental physics in the strong-field regime. In particular, these observations could possibly reveal a great deal of information on the nature of the central objects and especially, the existence or otherwise of event horizons in these objects.

Basically, what the EHT has observed in both Sgr A$^*$ and M87$^*$ and reported so far are images containing a shadow (a darker region over a brighter background) in some detail. Such an image can be formed if the central object is a black hole which necessarily has an event horizon and also therefore unstable photon orbits known as light rings (or a photon sphere in a spherically symmetric static case) around it. Light rings cause photons to undergo unboundedly large amount of bending (strong gravitational lensing) \citep{SL1,SL2,SL3}. They are also known as photon capture orbits as a slight perturbation on photons on such an orbit can cause them to be either captured by the black hole or sent off to a faraway observer. Therefore, the event horizon of a black hole, together with the unstable light rings, is expected to cast a characteristic shadow-like image of the photons emitted from nearby light sources or of the radiation emitted from an accretion flow around it \citep{shadow1,shadow2,shadow3}.

The gravitational field around astrophysical compact objects is believed to be described by the Kerr geometry which is a stationary, axially symmetric vacuum solution in general relativity (GR). The recent observation of Sgr A$^*$ by the EHT finds that the observed image is consistent with the expected appearance of a Kerr black hole \citep{SgrA_EHT1}. However, various non-vacuum solutions in GR as well as those in various modified gravity theories deviate from the Kerr (or Schwarzschild) solution and might potentially mimic many of the properties of the Kerr (or Schwarzschild) black hole \citep{mimicker1,mimicker2,mimicker3,mimicker4,mimicker5,mimicker5a,mimicker6,mimicker7,mimicker8, mimicker9,mimicker10,mimicker11}. Hence, in this EHT era, it becomes important to study images and shadows of these objects and compare them with the EHT results. This could help us constraining the parameter of these spacetimes (see e.g. \citep{SgrA_EHT2,mimicker3,Younsi,Mizuno,Olivares,test1,test2,test3, test4,test5,test6,test7,test8,test9,test10,test11,test12,test13,test14,test15,test16,test17}). In this regard, while the notion of black hole event horizons continues to attract much of the attention, horizonless compact objects which mimic black holes are also fast gaining popularity in the literature (see \cite{HCO1} and references therein). Among these potential black hole mimicker are wormholes and naked singularities (See e.g. \cite{mimicker1,mimicker2,mimicker3}).

In this work, we consider two black hole mimickers and constrain them using the EHT results of Sgr A$^*$. The first mimicker is a rotating generalisation of the recently proposed static, spherically symmetric black-bounce spacetime by Simpson and Visser (SV) where the central Schwarzschild singularity is replaced by a minimal surface of radius $r_0$ \citep{SV}. The SV metric is attractive in the sense that it is a minimal one parameter extension of the Schwarzschild metric and can describe a regular black hole or a wormhole for different choices of the parameter value. Various aspects of this metric and its rotating generalisation have been studied recently \citep{mimicker3,SV1,SV2,SV3,SV4,SV5,SV6,SV7,SV8,SV9,SV10,SV11,SV12,SV13,SV14}. The second mimicker is the Zipoy-Voorhees spacetime \citep{gamma1,gamma2,gamma3,gamma4,gamma5,gamma6} whose metric is popularly known as the $\gamma$-metric. It is a static, axially symmetric vacuum solution of Einstein's equations in GR. It is a singular solution without an event horizon and is characterized by two parameters $m$ and $\gamma$, where $m$ is related to the mass, and $\gamma$ measures deformation from the Schwarzschild solution. Various properties of this spacetime such as the global structure, shadows, strong lensing, motion of test particles, accretion disk properties etc. have been studied in the literature\citep{mimicker2, mimicker4,gamma5,gamma6,gamma7,gamma8,gamma9,gamma10,gamma11,gamma12,gamma13,gamma14,gamma15}. Our aim in this paper is to study shadows of these objects and constrain the parameters $r_0$ and $\gamma$ using Sgr A$^*$ observation.

This paper is organized as follows. In Sec. \ref{sec:metric}, we discuss the spacetime geometries of the black hole mimickers. In Sec. \ref{sec:shadows}, we study shadows of these objects. We constrain the parameters of these mimickers using the Sgr A$^*$ results in Sec. \ref{sec:results}. We conclude in Sec. \ref{sec:conclusion} with a brief summary of our results.

\section{The black hole mimickers and their spacetime geometries}
\label{sec:metric}

We start with the Kerr black hole metric which, in Boyer-Lindquist coordinates, can be written as
\begin{equation} 
ds^2=-\left(1-\frac{2Mr}{\Sigma}\right)dt^2-\frac{4Mar \sin^2\theta}{\Sigma}dt d\phi +\frac{\Sigma}{\Delta}dr^2 +\Sigma d\theta^2+\left( r^2+a^2+\frac{2Ma^2r \sin^2\theta}
{\Sigma}\right) \sin^2\theta d\phi^2,
\label{kerr}
\end{equation}
where $M$ is the mass of the black hole, $a$ is the specific angular momentum defined as $a=J/M$ and
\begin{equation}
 \Sigma=r^2+a^2 \cos^2\theta, \hspace{0.5cm}   \Delta=r^2-2Mr+a^2 .
\end{equation}
The black hole has event horizons at $r_\pm = M\pm\sqrt{M^2-a^2}$, where the `$+$' and `$-$' sign respectively correspond to the outer and inner horizon. We now briefly discuss below the geometries of the black hole mimicker which we are going to work with.

\subsection{Rotating Simpson-Visser (SV) metric: a special case of Johannsen metric}

Recently, Simpson and Visser constructed a regular spacetime metric by replacing the central Schwarzschild singularity by a non-singular minimal surface of radius $r_0$.  The spacetime geometry can acts as a black hole mimicker. It is given by \citep{SV}
\begin{equation}
ds^2=-\left(1-\frac{2M}{\sqrt{r^2+r_0^2}}\right)dt^2+\frac{dr^2}{1-\frac{2M}{\sqrt{r^2+r_0^2}}}+(r^2+r_0^2)(d\theta^2+\sin^2\theta d\phi^2).
\end{equation}
The above geometry represents a wormhole when $r_0>2M$ with $r_0$ being the wormhole throat radius. In the above coordinate system, the wormhole throat corresponds to $r=0$. However, when $r_0<2M$, the throat is hidden inside the event horizon $r_H=\sqrt{4M^2-r_0^2}$, thereby representing a regular black hole. A rotating version of the above spacetime has been constructed recently using Newman-Janis algorithm \citep{SV1} (see also \cite{mimicker3}). The rotating spacetime geometry, after using the coordinate transformation $\bar{r}=\sqrt{r^2+r_0^2}$ and dropping the bar, is given by \citep{mimicker3}
\begin{eqnarray} 
ds^2 &=&-\left(1-\frac{2Mr}{\Sigma}\right)dt^2-\frac{4Mar \sin^2\theta}{\Sigma}dt d\phi +\frac{\Sigma}{\Delta\hat{\Delta}}dr^2 \nonumber \\
& & +\Sigma d\theta^2 + \left( r^2+a^2+\frac{2Ma^2r \sin^2\theta}
{\Sigma}\right) \sin^2\theta d\phi^2,
\label{kerrWH}
\end{eqnarray}
\begin{equation}
 \Sigma=r^2+a^2 \cos^2\theta, \quad   \Delta=r^2-2Mr+a^2, \quad \hat{\Delta}=1-\frac{r_0^2}{r^2}.
\end{equation}
Note that the above spacetime is very similar to the Kerr geometry and differs from the Kerr only by the term $\hat{\Delta}$. For $r_0=0$, $\hat{\Delta}=1$ and it reduces the Kerr black hole metric with the event horizons given by $r_\pm$. The above metric also has horizons at $r_{\pm}$. However, depending on the relative values of $r_0$ and the spin $a$, the horizons may or may not be relevant. For $0\leq a/M\leq 1$, the above geometry represents a black hole when $r_0<r_{+}$ and a wormhole when $r_0\geq r_{+}$. However, for $a/M>1$, $r_{\pm}$ do not exist, and hence, it always represents a wormhole with the throat given by $\hat{\Delta}=0$, i.e., by $r=r_0$.

It can be shown that the above rotating SV metric can be obtained as a special case of the parametrized non-Kerr metric given by Johannsen. The non-Kerr metric by Johannsen \citep{johannsen} is given by
\begin{eqnarray}
g_{tt} &=& -\frac{\tilde{\Sigma}[\Delta-a^2A_2(r)^2\sin^2\theta]}{[(r^2+a^2)A_1(r)-a^2A_2(r)\sin^2\theta]^2}, \nonumber \\
g_{t\phi} &=& -\frac{a[(r^2+a^2)A_1(r)A_2(r)-\Delta]\tilde{\Sigma}\sin^2\theta}{[(r^2+a^2)A_1(r)-a^2A_2(r)\sin^2\theta]^2}, \nonumber \\
g_{rr} &=& \frac{\tilde{\Sigma}}{\Delta A_5(r)}, \nonumber \\
g_{\theta \theta} &=& \tilde{\Sigma}, \nonumber \\
g_{\phi \phi} &=& \frac{\tilde{\Sigma} \sin^2 \theta \left[(r^2 + a^2)^2 A_1(r)^2 - a^2 \Delta \sin^2 \theta \right]}{[(r^2+a^2)A_1(r)-a^2A_2(r)\sin^2\theta]^2},
\label{eq:metric}
\end{eqnarray}
where
\begin{eqnarray}
A_1(r) &=& 1 + \sum_{n=3}^\infty \alpha_{1n} \left( \frac{M}{r} \right)^n, 
\label{eq:A1}\\
A_2(r) &=& 1 + \sum_{n=2}^\infty \alpha_{2n} \left( \frac{M}{r} \right)^n, 
\label{eq:A2}\\
A_5(r) &=& 1 + \sum_{n=2}^\infty \alpha_{5n} \left( \frac{M}{r} \right)^n, 
\label{eq:A5}\\
\tilde{\Sigma} &=& \Sigma + f(r), 
\label{eq:Sigmatilde}\\
f(r) &=& \sum_{n=3}^\infty\epsilon_n \frac{M^n}{r^{n-2}}.
\label{eq:f}
\end{eqnarray}
The Kerr metric is recovered when $f(r)=0$, $A_1(r)=A_2(r)=A_5(r)=1$, i.e., $\alpha_{1n}=\alpha_{2n}=\alpha_{5n}=\epsilon_{n}=0$. The rotating SV metric in Eq. (\ref{kerrWH}) can be obtained for the special case $f(r)=0$, $A_1(r)=A_2(r)=1$ and $A_5(r)=\hat{\Delta}(r)$, i.e., for
\begin{equation}
\alpha_{1n}=\alpha_{2n}=\epsilon_{n}=0, \quad \alpha_{5n}=0\;(n\neq 2), \quad \alpha_{52}=-\frac{r_0^2}{M^2}.
\end{equation}

\subsection{The $\gamma$-metric}

The Zipoy-Voorhees spacetime \citep{gamma1,gamma2,gamma3,gamma4,gamma5,gamma6} whose metric is popularly known as the $\gamma$-metric is a static, axially-symmetric vacuum solution of Einstein's equations in GR. The metric belongs to a Weyl's class, and in Erez-Rosen coordinates, is given by 
\begin{align}
ds^2 = -A(r)dt^2 + \frac{1}{A(r)}\left[ B(r,\theta)dr^2 + C(r,\theta) d\theta^2 + (r^2-2mr)\sin^2\theta d\phi^2 \right],
\label{eq:metric_gamma}
\end{align}
where the functions $ A,B,C $ are given as
\begin{equation}
A(r) = \left( 1-\frac{2m}{r} \right)^\gamma,~B(r,\theta) = \left( \frac{r^2-2mr}{r^2-2mr+m^2\sin^2\theta} \right)^{\gamma^2-1},~
C(r,\theta) = \frac{(r^2-2mr)^{\gamma^2}}{(r^2-2mr+m^2\sin^2\theta)^{\gamma^2-1}}.
\end{equation}

The metric is characterized by two parameters $m$ and $\gamma$, $m$ being related to the mass 
and $\gamma$ measuring deformation of the spacetime from spherical symmetry. The 
Arnowitt-Deser-Misner (ADM) mass of the spacetime is $ M=m\gamma $, and the corresponding 
quadrupole moment is $Q=\frac{\gamma M^3}{3}(1-\gamma^2)$ \citep{gamma5}. The monopole $M$ 
and the quadrupole $Q$ are the only independent components of multipole moments, as all higher 
order components can be expressed in terms of $M$ and $Q$. The spacetime metric reduces to the flat Minkowski spacetime for $\gamma=0$ and to the spherically symmetric 
Schwarzschild black hole solution for $\gamma=1$. For all other values of $\gamma$, it has a curvature singularity at $r=2m=2M/\gamma$ which is not covered inside any event horizon and is deformed from spherically symmetric 
to axially symmetric, with $\gamma<1$ $(\gamma>1)$ representing a prolate (oblate) spheroid. The singularity at $r=2m$ (Schwarzschild horizon for $\gamma=1$) represents an infinitely red-shifted surface and can thus exhibit observational properties analogous to the event horizon of a black hole for observers at infinity \citep{mimicker2}.

We consider the spacetime metric in limiting case $\gamma \to \infty$ also, keeping the ADM 
mass $M$ fixed and finite. The resulting spacetime corresponds to the Chazy-Curzon solution of GR 
\citep{gamma4,gamma16,gamma17,gamma18}. We call it as 
Gamma-Infinite (GI) spacetime for simplicity. The corresponding spacetime is given by
\begin{equation}
ds^2 = -\exp\left(-\frac{2M}{r}\right)dt^2 + \exp\left(\frac{2M}{r}\right)\left[\exp\left(-\frac{M^2\sin^2\theta}{r^2}\right)
\left(dr^2 + r^2 d\theta^2\right) + r^2\sin^2\theta d\phi^2\right].
\label{eq:metric_GI}
\end{equation}

\section{Shadows of the black hole mimickers}
\label{sec:shadows}

We first begin with the shadow of a Kerr black hole as it will help us to understand the same for the rotating SV metric. The separated null geodesic equations which are required for the purpose of calculating shadows are given by
\begin{equation}
\Sigma\frac{dr}{d\lambda}=\pm \sqrt{R(r)},
\label{eq:r_eqn}
\end{equation}
\begin{equation}
\Sigma\frac{d\theta}{d\lambda}=\pm \sqrt{\Theta(\theta)},
\label{eq:theta_eqn}
\end{equation}
where
\begin{equation}
R(r)=\left[(r^2+a^2)E-aL\right]^2-\Delta\left[\mathcal{K}+\left(L-aE\right)^2\right],
\end{equation}
\begin{equation}
\Theta(\theta)=\mathcal{K}+a^2E^2\cos^2\theta-L^2\cot^2\theta,
\end{equation}
$E$ is the energy of photons, $L$ is the angular momentum about the axis of symmetry, and $\mathcal{K}$ is the carter constant. The unstable photon orbits, which form the boundary of a shadow, are given by $\dot{r}=0$, $\ddot{r}=0$ and $\dddot{r}>0$ which, in terms of $R(r)$, become
\begin{equation}
R(r_{ph})=0,\quad R'(r_{ph})=0, \quad R''(r_{ph})>0,
\label{eq:R_condition}
\end{equation}
where $r_{ph}$ is the radius of an unstable photon orbit. From the first two conditions, we obtain the critical impact parameters
\begin{equation}
\xi_{ph}=\frac{4Mr_{ph}^2-(r_{ph}+M)(r_{ph}^2+a^2)}{a(r_{ph}-M)},
\label{eq:xi}
\end{equation}
\begin{equation}
\eta_{ph}=\frac{4Ma^2r_{ph}^3-r_{ph}^2\left[r_{ph}(r_{ph}-3M)\right]^2}{a^2(r_{ph}-M)^2},
\label{eq:eta}
\end{equation}
where $\xi=L/E$ and $\eta=\mathcal{K}/E^2$. However, the apparent shape of a shadow in a observer's sky is described using the celestial coordinates defined by \citep{celestial}
\begin{equation}
\alpha=\lim_{r_o\to\infty}\left(-r_o^2\sin\theta_o\frac{d\phi}{dr}\Big\vert_{(r_o,\theta_o)}\right)=-\frac{\xi}{\sin\theta_o},
\label{eq:alpha}
\end{equation}
\begin{equation}
\beta=\lim_{r_o\to\infty}\left(r_o^2\frac{d\theta}{dr}\Big\vert_{(r_o,\theta_o)}\right)=\pm \sqrt{\eta+a^2\cos^2\theta_o-\xi^2\cot^2\theta_o},
\label{eq:beta}
\end{equation}
where we have used the separated null geodesic equations in the last steps, and $(r_o,\theta_o)$ are the position coordinates of the asymptotic observer. The contour of the shadow in the observer's sky ($\alpha-\beta$ plane) is obtained through a parametric plot of $\alpha_{ph}$ and $\beta_{ph}$ where
\begin{equation}
\alpha_{ph}=-\frac{\xi_{ph}}{\sin\theta_o},
\label{eq:alpha_ph}
\end{equation}
\begin{equation}
\beta_{ph}=\pm \sqrt{\eta_{ph}+a^2\cos^2\theta_o-\xi_{ph}^2\cot^2\theta_o}.
\label{eq:beta_ph}
\end{equation}

\begin{figure}[]
\centering
\includegraphics[scale=0.9]{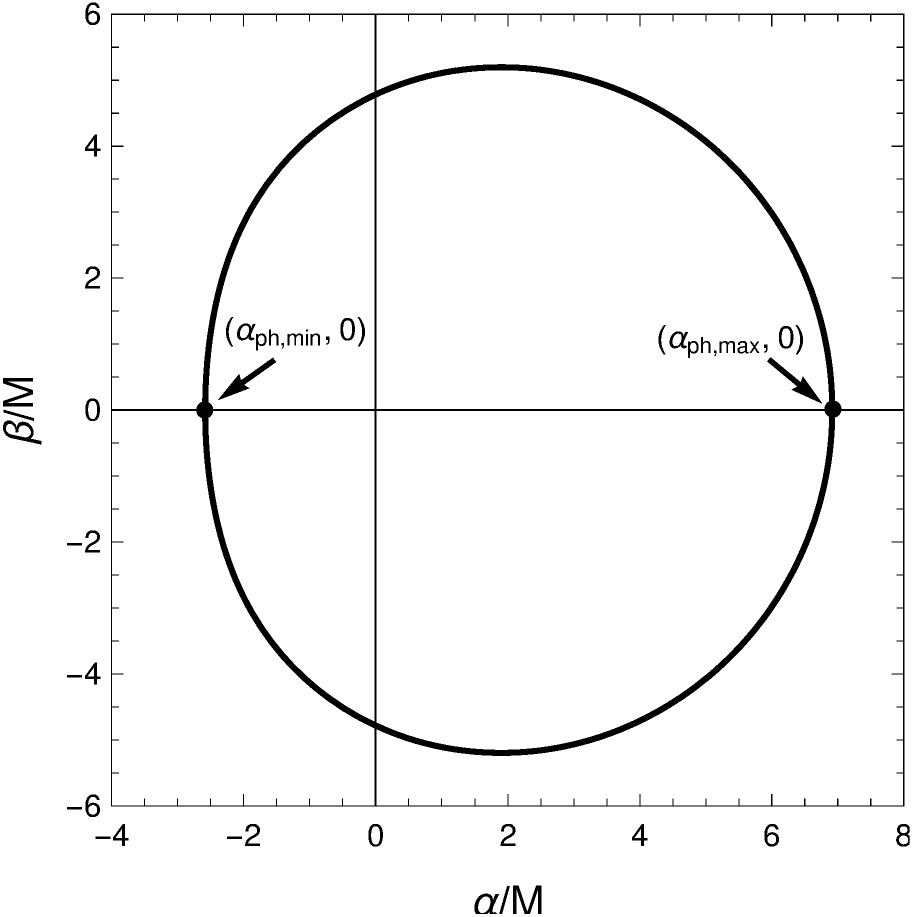}
\caption{The shadow of a Kerr black hole for $a/M=0.95$ and $\theta_o=90^\circ$.}
\label{fig:shadowKerr}
\end{figure}

The unstable photon orbits in a Kerr spacetime geometry consist of both prograde (orbits on which photons have motion in the direction of the spin) and retrograde (orbits on which photons have motion in a direction opposite to the spin) orbits. The prograde orbits have lesser radii than those of the retrograde ones. A typical shadow of a Kerr black hole is shown in Fig. \ref{fig:shadowKerr}. The spin axis, when projected on the $\alpha-\beta$ plane, coincides with the $\beta$-axis and divides the shadow into two parts. The part of the shadow contour on the left side (i.e., on $\alpha<0$ side) of the spin axis is due to the unstable photon orbits which are prograde ($\xi>0$, i.e., $L>0$), while the one on the other side (i.e., on $\alpha>0$ side) is due to the unstable photon orbits which are retrograde ($\xi<0$, i.e., $L<0$). Among the unstable photon orbits which form the shadow, the one which has the minimum radius $r_{ph,min}$ is of prograde type and corresponds to the point $(\alpha_{ph,min},0)$ on the shadow contour. On the other hand, the one which has the maximum radius $r_{ph,max}$ is of retrograde type and corresponds to the point $(\alpha_{ph,max},0)$ on the shadow contour. Therefore, the shadow is formed by the unstable photon orbits whose radii are in the range $r_{ph,min}\leq r_{ph}\leq r_{ph,max}$. The minimum and maximum radii $r_{ph,min}$ and $r_{ph,max}$ are obtained from $\beta_{ph}=0$, i.e., from
\begin{equation}
\eta_{ph}+a^2\cos^2\theta_o-\xi_{ph}^2\cot^2\theta_o=0.
\end{equation}
$r_{ph,max}$ and $r_{ph,min}$ ($\leq r_{ph,max}$) are respectively given by the largest and the second largest roots of the above equation. Note that, apart from the mass $M$ and the spin $a$ of the black hole, $r_{ph,max}$ and $r_{ph,min}$ depend on the observation (inclination) angle $\theta_o$.

\subsection{Rotating SV metric}
\label{sec:shadows_RSV}

The method of calculating shadow for this metric has been discussed in great details in \cite{mimicker3}. However, we again briefly discuss it here for convenience. The separated null geodesic equations in this case are the same as those in the Kerr black hole spacetime except that the radial equation (\ref{eq:r_eqn}) is modified to
\begin{equation}
\Sigma\frac{dr}{d\lambda}=\pm\sqrt{\hat{\Delta}(r)} \sqrt{R(r)},
\end{equation}
where $R(r)$ is the same as that in the Kerr black hole case.

For unstable photon orbits which completely lie outside of the surface $r_0$ (i.e. for $r_{ph}> r_0$), the unstable orbit conditions $\dot{r}=0$, $\ddot{r}=0$ and $\dddot{r}>0$ reduce to $R(r_{ph})=0$, $R'(r_{ph})=0$ and $R''(r_{ph})>0$ which are the same as those in the Kerr black hole case. Therefore, the part of the shadow contour which is formed by these unstable orbits is given by Eqs. (\ref{eq:alpha_ph}) and (\ref{eq:beta_ph}). However, in a wormhole case, it is known that the wormhole throat can also acts as a natural location of unstable or stable photon orbits (see \cite{rajibul_2018,rajibul_2019} for details). Therefore, for unstable photon orbits which lie at the throat (i.e. for $r_{ph}=r_0$), $\dot{r}(r_{ph})=0$ is satisfied automatically as $\hat{\Delta}(r_{ph})=0$. Therefore, for such orbits, the unstable orbit conditions $\dot{r}=0$, $\ddot{r}=0$ and $\dddot{r}>0$ now reduce to $R(r_0)=0$ and $R'(r_{0})>0$, which are different from the ones for the unstable orbits lying outside the throat. Now, $R(r_0)=0$ gives
\begin{equation}
\left[(r_0^2+a^2)-a\xi_0\right]^2-\Delta(r_0)\left[\eta_0+\left(\xi_0-a\right)^2\right]=0,
\label{eq:xi-eta-r0}
\end{equation}
where $\xi_0$ and $\eta_0$ denote the critical impact parameters of photons comprising the unstable orbits at the throat $r_0$. After using Eqs. (\ref{eq:alpha}) and (\ref{eq:beta}) in the last equation, we obtain
\begin{equation}
\left[(r_0^2+a^2)+a\sin\theta_o\alpha_0\right]^2-\Delta(r_0)\left[\beta_0^2+\left(\alpha_0+a\sin\theta_o\right)^2\right]=0.
\label{eq:alpha-beta-0}
\end{equation}
The above equation gives the part of the shadow contour which is formed by the unstable photon orbits located at the throat.

Now, depending on the relative values of $r_0$, $r_{ph,min}$ and $r_{ph,max}$, we have following three cases here.

\noindent {\bf I. $r_0\leq r_{ph,min}$:} The shadow in this case is completely given by Eqs. (\ref{eq:alpha_ph}) and (\ref{eq:beta_ph}) and is the same as that of the Kerr black hole. Therefore, this case perfectly mimic the Kerr black hole.

\noindent {\bf II. $r_{ph,min}<r_0\leq r_{ph,max}$:} In this case, the unstable photon orbits having radii in the range $r_{ph,min}\leq r_{ph}<r_0$ become irrelevant as they are smaller than the allowed minimum value $r_0$ of the radial coordinate. As a results, a part of the shadow contour given by Eqs. (\ref{eq:alpha_ph}) and (\ref{eq:beta_ph}) is lost due to these unstable photon orbits which become irrelevant. This lost part is now compensated by the unstable photon orbits which lie at the throat $r_0$. Therefore, in this case, the complete shadow contour is given by the union of the $(\alpha_{ph},\beta_{ph})$ curve [Eqs. (\ref{eq:alpha_ph}) and (\ref{eq:beta_ph})] with $r_0\leq r_{ph}\leq r_{ph,max}$ and the $(\alpha_0,\beta_0)$ curve [Eq. (\ref{eq:alpha-beta-0})]. Note that, in order to have a closed shadow contour, the two curves must intersect at two points, say at $(\alpha_{0,max},\pm \beta_{0,max})$. As the intersection points must correspond to $r_{ph}=r_0$, we have $\alpha_{0,max}=\alpha_{ph}\big\vert_{r_{ph}=r_0}$ and $\beta_{0,max}=\beta_{ph}\big\vert_{r_{ph}=r_0}$. Therefore, the ($\alpha_0,\beta_0$) curve in this case has the ranges $\alpha_{0,min}\leq \alpha_0\leq \alpha_{0,max}$ and $-\beta_{0,max}\leq \beta_0\leq \beta_{0,max}$, where $\alpha_{0,min}$ is obtained by putting $\beta_0=0$ in Eq. (\ref{eq:alpha-beta-0}). This gives
\begin{equation}
\alpha_{0,min}=\frac{r_0^2+a^2\mp a\sin\theta_o\sqrt{\Delta(r_0)}}{\pm\sqrt{\Delta(r_0)}-a\sin\theta_o},
\end{equation}
where we take the root which is negative and close to $\alpha_{0,max}$.

\noindent {\bf III. $r_0> r_{ph,max}$:} In this case, all the unstable photon orbits given by the conditions $R(r_{ph})=0$, $R'(r_{ph})=0$ and $R''(r_{ph})>0$ become irrelevant as their radii are smaller than the allowed minimum value $r_0$ of the radial coordinate. In such a case, the shadow contour is completely given by the $(\alpha_0,\beta_0)$ curve [Eq. (\ref{eq:alpha-beta-0})].

\begin{figure}[h!]
\centering
\includegraphics[scale=0.55]{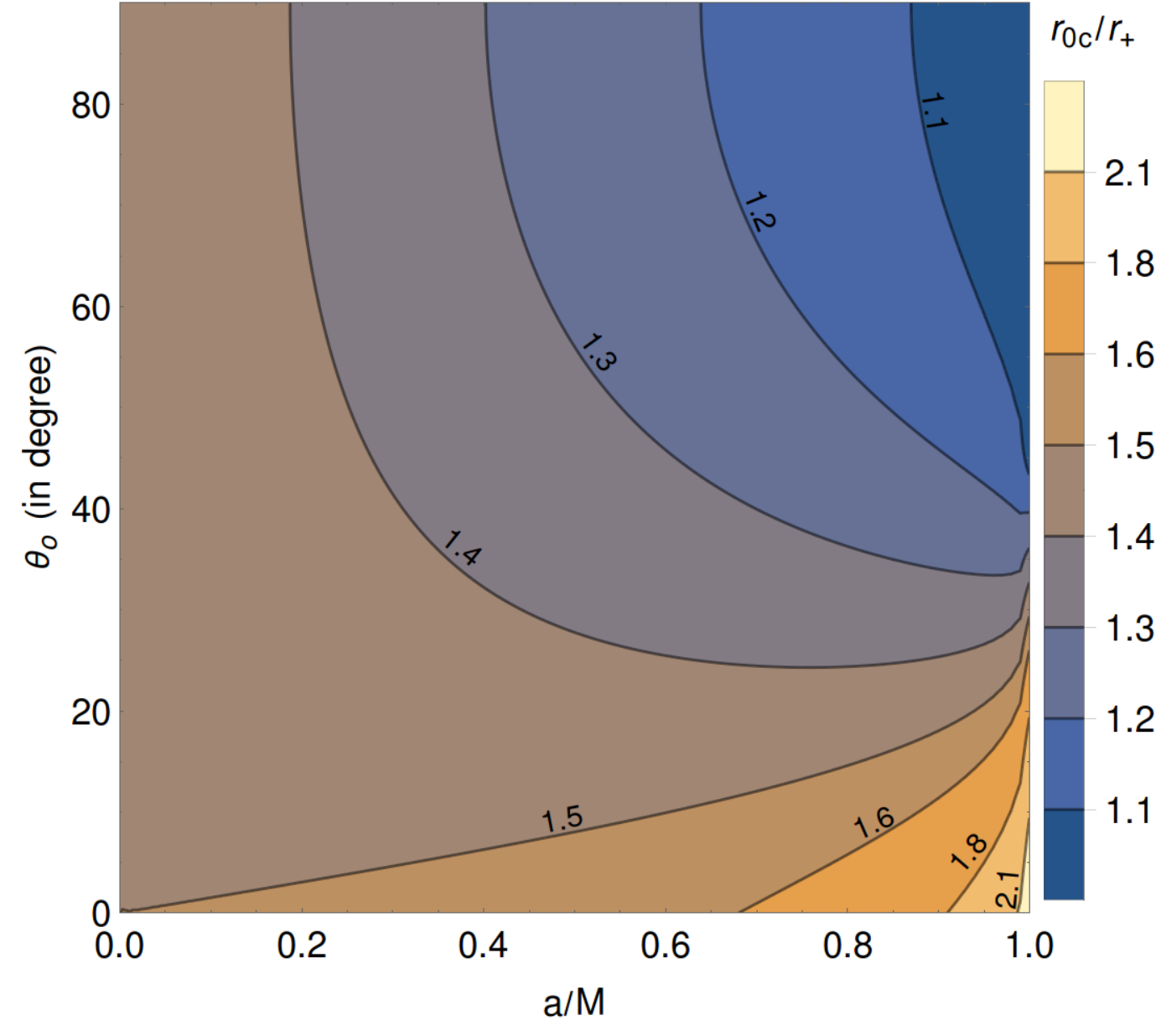}
\caption{Plot showing the critical value $r_{0c}$ above which the shadow of the rotating SV metric starts deviating from that of the Kerr black hole. This plot has also been considered in \cite{mimicker3}.}
\label{fig:r0c}
\end{figure}

\begin{figure}[]
\centering
\subfigure{\includegraphics[scale=0.65]{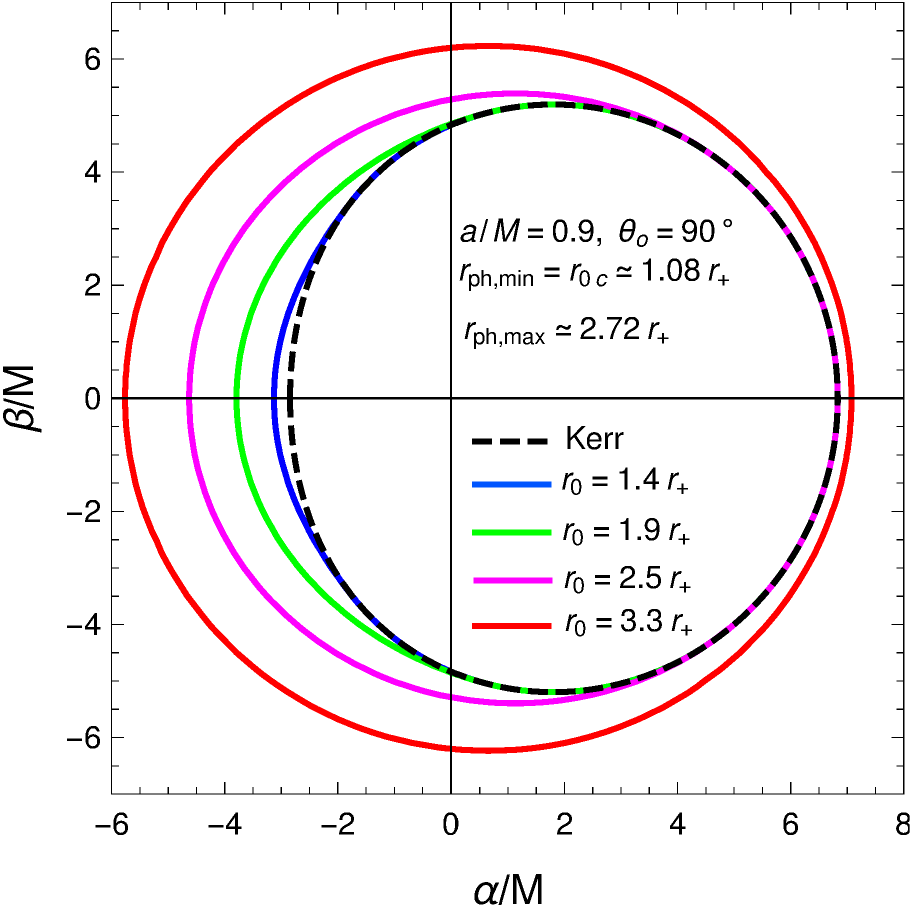}}
\subfigure{\includegraphics[scale=0.65]{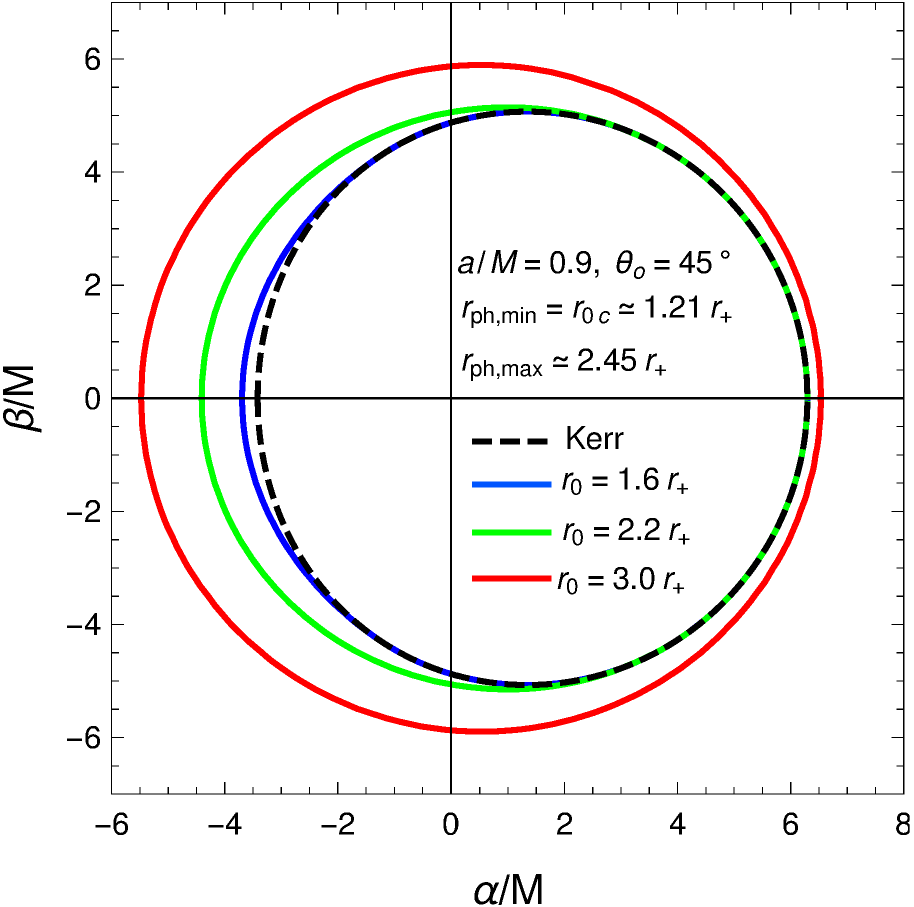}}
\subfigure{\includegraphics[scale=0.65]{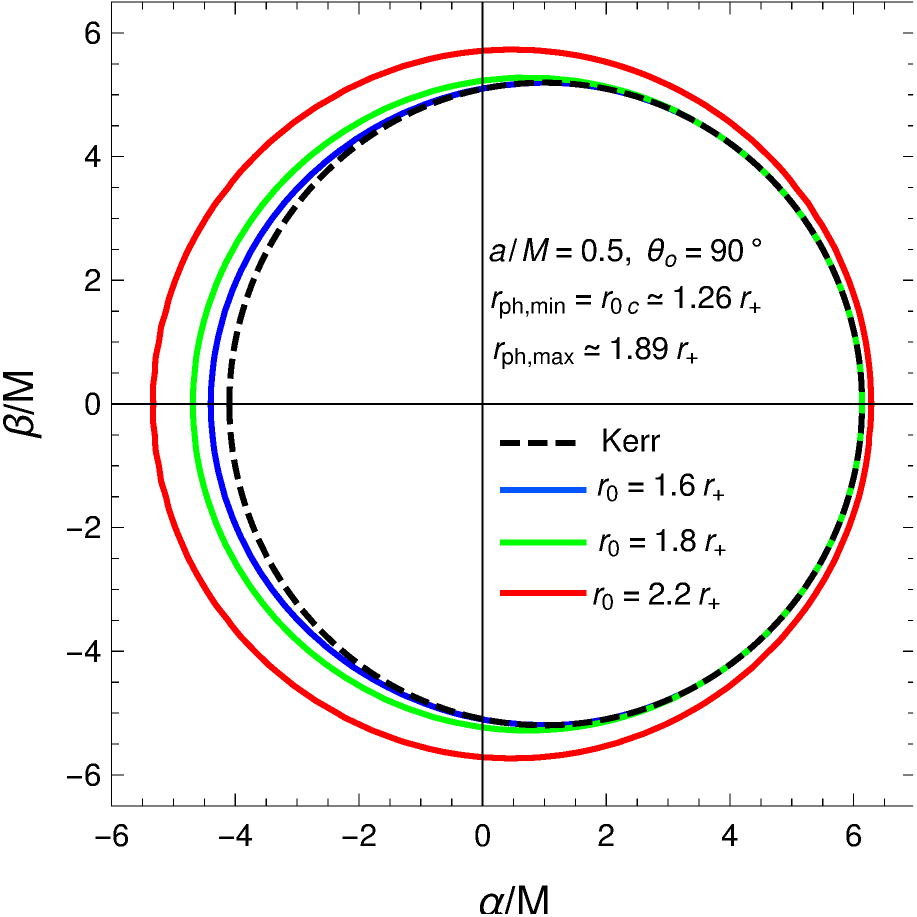}}
\subfigure{\includegraphics[scale=0.65]{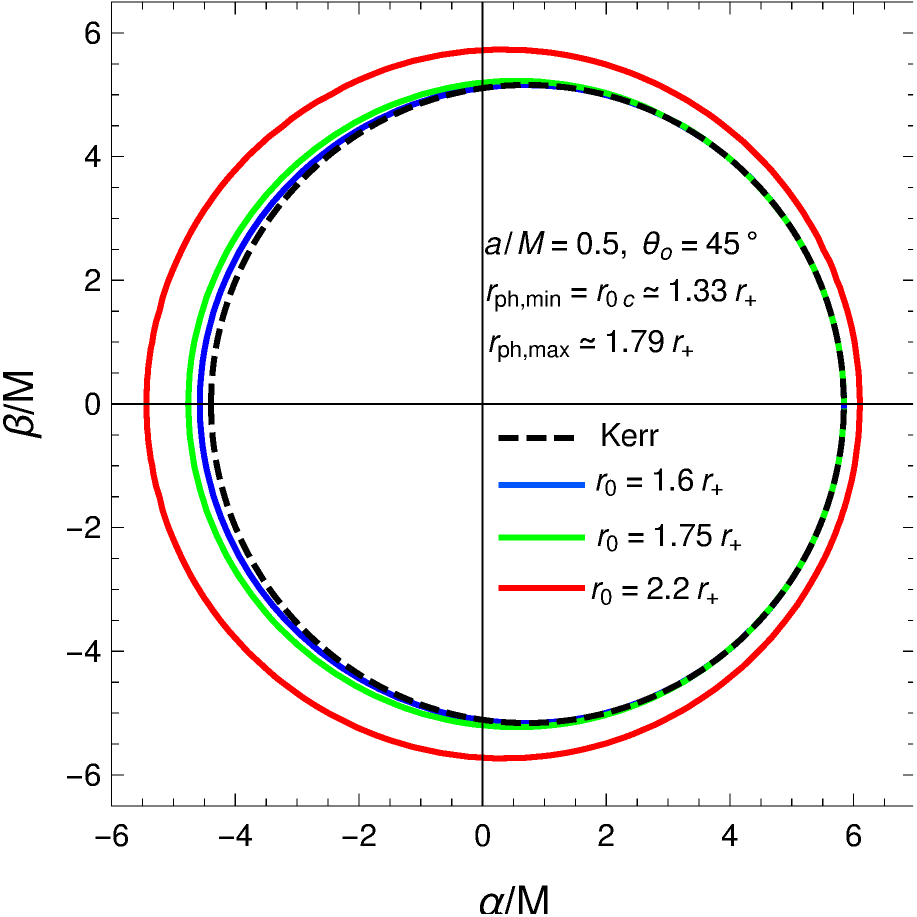}}
\caption{Shadow of the rotating SV metric for different values of $r_0$, spin $a$ and inclination angle $\theta_o$. Note that, for $r_0\leq r_{0c}$, the shadow is the same as that of the Kerr black hole. It only starts deviating from that of the Kerr black hole for $r_0>r_{0c}$.}
\label{fig:RSV_shadow}
\end{figure}

A more detailed discussion of the above method of calculating shadow for the rotating SV metric is available in \cite{mimicker3}. It is to be noted that the shadow of the rotating SV metric deviates from that of the Kerr black hole only when $r_0>r_{ph,min}$. Therefore, we define a critical value of $r_0$ by $r_{0c}=r_{ph,min}$. For $r_0\leq r_{0c}$, the shadow is the same as that of the Kerr black hole. However, if  $r_0> r_{0c}$, then the shadow deviates from that of the Kerr black hole. Figure \ref{fig:r0c} shows the dependence of the critical radius $r_{0c}$ on the spin and observation angle. Figure \ref{fig:RSV_shadow} shows shadows of the rotating SV metric for different $r_0$, spin $a$ and inclination angle $\theta_o$. Note that, for a given $a$ and $\theta_o$, the shadow size is bigger than that of the Kerr black hole when $r_0>r_{0c}$ and increases with increasing $r_0$ ($>r_{0c}$). Note also that, for a given $r_0>r_{0c}$, the shadow seems to be more rounded than that of the Kerr black hole with the same $a$ and $\theta_o$.

\subsection{$\gamma$-metric}
\label{sec:shadows_gamma}

\begin{figure}[]
\centering
\subfigure{\includegraphics[scale=0.65]{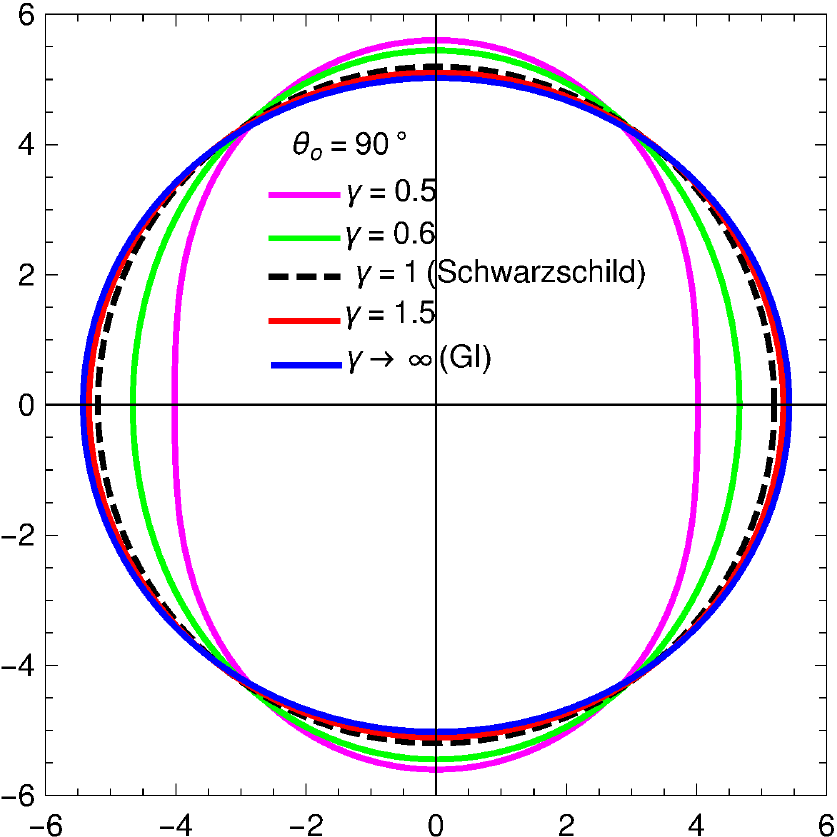}}
\subfigure{\includegraphics[scale=0.65]{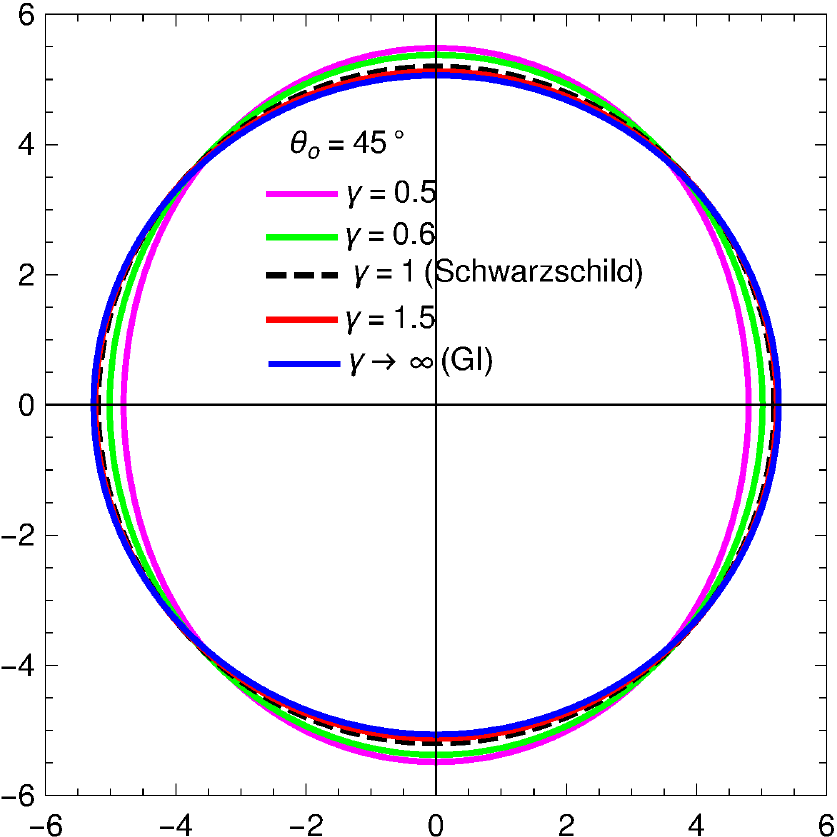}}
\subfigure{\includegraphics[scale=0.65]{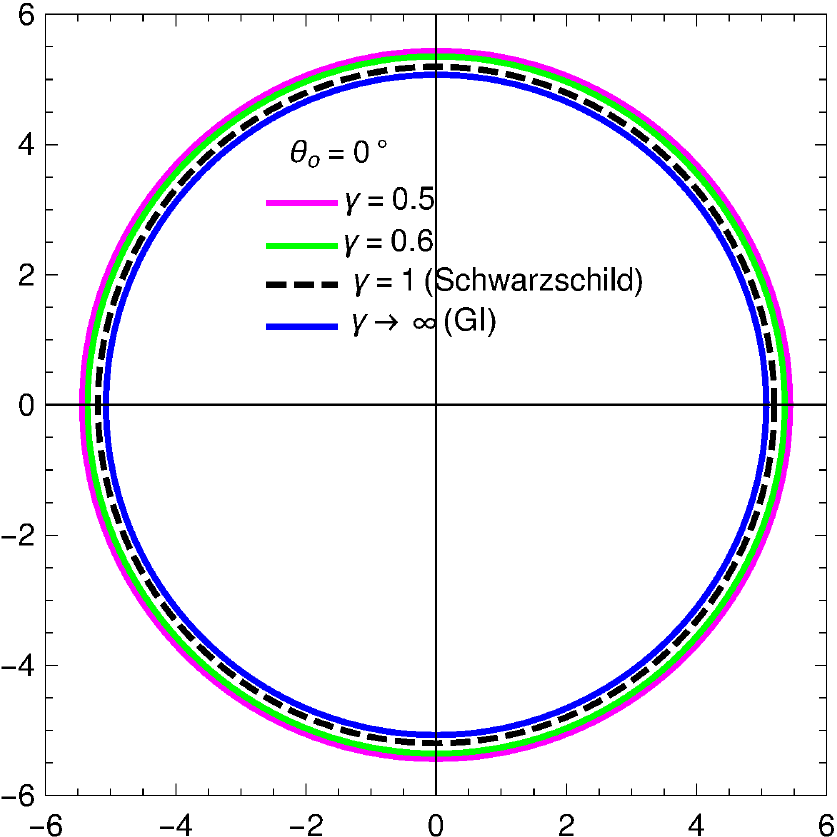}}
\caption{Shadow of the $\gamma$-metric for different values of $\gamma$ and inclination angle $\theta_o$.}
\label{fig:gamma_shadow}
\end{figure}

The shadow of this metric has been discussed in \cite{mimicker2,mimicker4}. This spacetime casts shadow for $\gamma\geq 1/2$. The null geodesic equations in this spacetime can not be separated. So, one can not obtain an analytic expression for the shadow in this case. We, therefore, use our numerical ray-tracing techniques used in some of our previous works \citep{mimicker4,mimicker9} and obtain the shadow. We do not discuss the details of the numerical techniques here as it can be found in the references just mentioned. Figure \ref{fig:gamma_shadow} shows the shadow of the $\gamma$-metric for different values of $\gamma$ and inclination angle $\theta_o$. It is to be noted that, with increasing $\gamma$ from $0.5$ to $\infty$, the change in the shadow shape is relatively slower for $\gamma>1$ than that for $\gamma<1$. Also, note that the shadow is of prolate and oblate shape for $\gamma<1$ and $\gamma>1$, respectively.

\section{Testing the black hole mimickers using Sgr A$^*$ observation}
\label{sec:results}

We now test and constrain the black mimickers using EHT observation of Sgr A$^*$. We use observables such as the average radius of the shadow and its deformation from circularity for this purpose. As the shadow has a reflection symmetry about the $\alpha$-axis on the $\alpha-\beta$ plane, its geometric center $(\alpha_c,\beta_c)$ lies on the $\alpha$-axis and is given by $\alpha_c=(1/A)\int \alpha dA$ and $\beta_c=0$, $dA$ being an area element. With this, we draw a vector $\overrightarrow{l}$ from the geometric center to a point $(\alpha,\beta)$ on the boundary of the shadow and define an angle $\phi$ between the $\alpha$-axis and the vector $\overrightarrow{l}$. We define the average radius $R_{sh}$ of the shadow as \citep{rav}
\begin{equation}
R_{sh}^2=\frac{1}{2\pi}\int_{0}^{2\pi}l^2(\phi)\; d\phi,
\end{equation}
where $l(\phi)=\sqrt{(\alpha(\phi)-\alpha_c)^2+\beta(\phi)^2}$ and $\phi=tan^{-1}(\beta(\phi)/(\alpha(\phi)-\alpha_c))$. Note that the above definition of the average radius $R_{sh}$ is equivalent to equating the total area of the shadow with the area of a circular disk of radius $R_{sh}$. One can also define the deviation of the shadow from circularity as
\begin{equation}
\Delta C=\frac{1}{R_{sh}}\sqrt{\frac{1}{2\pi}\int_{0}^{2\pi}(l(\phi)-R_{sh})^2\; d\phi}.
\end{equation}
Note that $\Delta C$ is the fractional RMS distance from the average radius. The average diameter of the shadow is given by $d_{sh}=2R_{sh}$.

The recent EHT papers on Sgr A$^*$ observation have used the fractional deviation parameter $\delta$ to constrain different spacetime geometries \citep{SgrA_EHT2}. $\delta$ quantifies the fractional deviation of the shadow diameter from that of a Schwarzschild black hole and is given by
\begin{equation}
\delta=\frac{d_{sh}}{d_{sh,Sch}}-1=\frac{R_{sh}}{3\sqrt{3}M}-1=\frac{\hat{R}_{sh}}{3\sqrt{3}}-1,
\end{equation}
where $\hat{R}_{sh}=R_{sh}/M$ is the dimensionless radius of the shadow and $d_{sh,Sch}=6\sqrt{3}M$ is the diameter of the Schwarzschild black hole shadow (in $G=C=1$ unit). Using the angular size of the observed shadow of Sgr A$^*$ and two separate set of priors on the mass-to-distance ratio from the Very Large Telescope Interferometer (VLTI) and Keck observations, EHT collaboration has provided the following bounds on the fractional deviation parameter $\delta$ \citep{SgrA_EHT1,SgrA_EHT2}
\begin{equation}
\delta=\left\{
  \begin{array}{ll}
  -0.08^{+0.09}_{-0.09}  & \;\;(\mbox{VLTI}) \\
  -0.04^{+0.09}_{-0.10}  & \;\;(\mbox{Keck})
  \end{array}
\right..
\end{equation}
Therefore, at the $1\sigma$ credible level, the fractional deviation parameter lies in the range $-0.17\leq \delta\leq 0.01$ (VLTI) and $-0.14\leq \delta\leq 0.05$ (Keck). We use these bounds to constrain the black hole mimickers under consideration. It is to be noted that, when we consider the VLTI and Keck bounds, the masses of the black hole mimickers are considered to be the same as those measured by the corresponding observations (VLTI and Keck).

\begin{figure}[h!]
\centering
\subfigure{\includegraphics[scale=0.42]{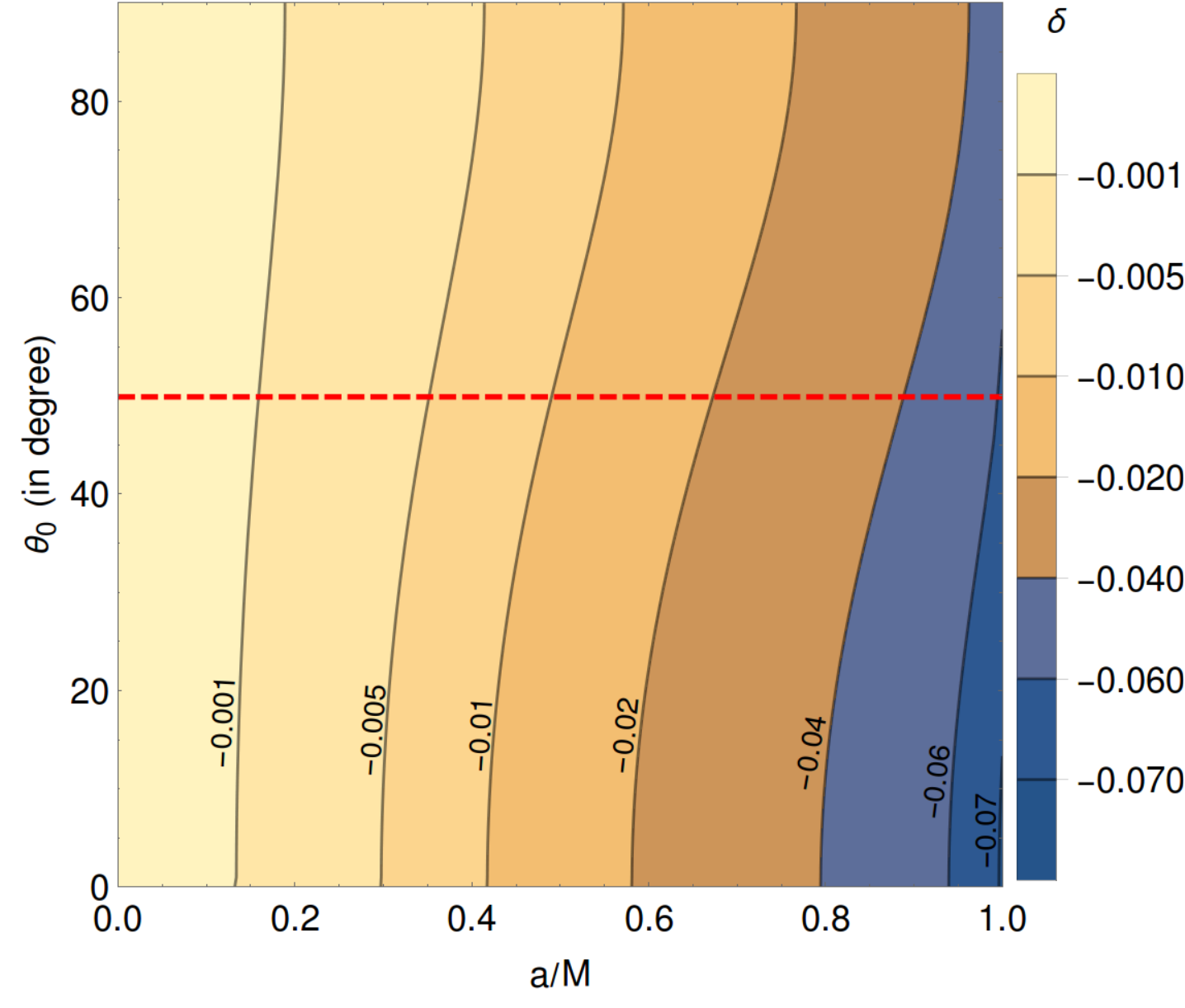}}
\subfigure{\includegraphics[scale=0.40]{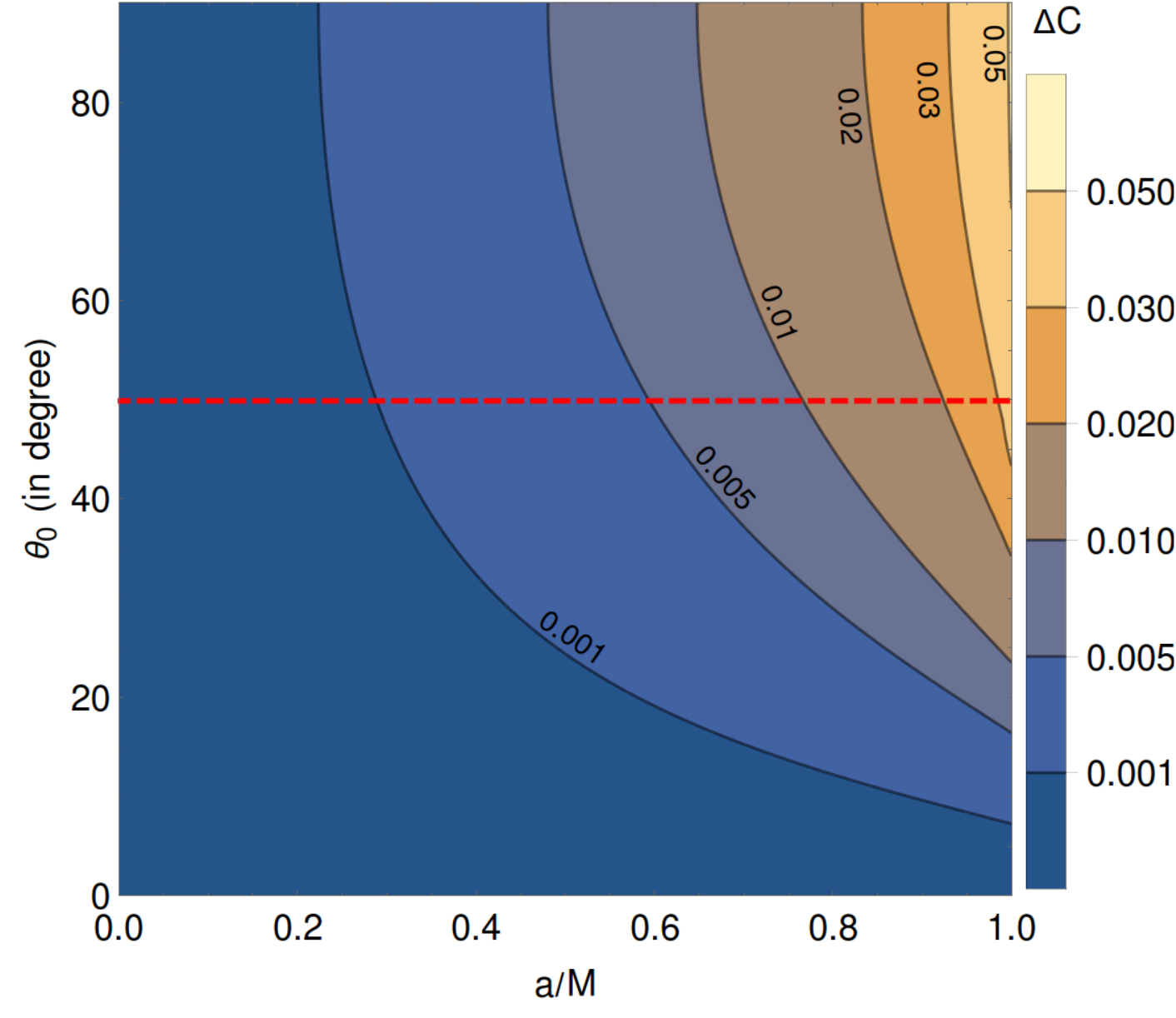}}
\caption{Plots showing the fractional deviation parameter $\delta$ and the deviation from circularity $\Delta C$ of the Kerr black hole shadow for different values of the spin and inclination angle. The red dashed curve shows the inclination angle $\theta_o=50^\circ$.}
\label{fig:Kerr_delta}
\end{figure}

Figure \ref{fig:Kerr_delta} shows the fractional deviation parameter $\delta$ and the deviation from circularity $\Delta C$ of the Kerr black hole shadow for different values of the spin and inclination angle. Note that $\delta\leq 0$ always for the Kerr black hole shadow. We have found, considering all the spins and inclination angles, that $-0.071\lesssim \delta \leq 0$, i.e., the Kerr shadow can be upto $7.1\%$ smaller than the Schwarzschild black hole shadow. Note that the Kerr black hole shadow is consistent with the observed shadow of Sgr A$^*$ as its $\delta$ lies within both the VLTI and Keck bounds. Although the EHT collaboration has not provided any constraint on the spin $a$ and found the non-rotating case to be disfavoured, they considered several non-rotating compact objects and constrain them using their results \citep{SgrA_EHT1}. They found that an inclination angle higher than $50^\circ$ (i.e., $\theta_o>50^\circ$) is disfavoured \citep{SgrA_EHT1}. Also, they found some promising models with $< 30^\circ$ inclination angle while constraining the inclination and black hole spin from the theoretical model comparison \citep{SgrA_EHT3}. Moreover, GRAVITY Collaboration have provided the spin to be $134^\circ$ (or equivalently $46^\circ$) \citep{Gravity_2019}, although the possibility of even low inclination angle from the observation of hotspot motion has also been discussed \citep{Gravity_2018}. Therefore, although we obtain our results for all possible inclination angle below, we mainly focus on $\theta_o\leq 50^\circ$. The deviation from circularity $\Delta C$ lies in the range $0\leq \Delta C\lesssim 0.0534$, i.e., the maximum deviation of the Kerr black hole shadow from circularity is about $5.34\%$ (See Fig. \ref{fig:Kerr_delta}). However, if we restrict the inclination angle to $50^\circ$ or below, i.e., if $\theta_o\leq 50^\circ$, then $0\leq \Delta C\lesssim 0.0374$, i.e., the maximum deviation from circularity within this inclination angle is about $3.74\%$, considering all spins.

We now constrain the black hole mimickers under consideration.

\subsection{Rotating SV metric}

As we have discussed in Subsec. \ref{sec:shadows_RSV}, the shadow of the rotating SV metric is the same as that of the Kerr black hole for $r_0\leq r_{0c}$. Therefore, for $r_0\leq r_{0c}$, the fractional deviation $\delta$ of the shadow is the same as that of the Kerr black hole ($\delta_{Kerr}$) with the same spin and inclination angle, i.e., $\delta=\delta_{Kerr}$ for $r_0\leq r_{0c}$ and hence, satisfies both VLTI and Keck bounds. The shadow deviates from that of the Kerr black hole when $r_0>r_{0c}$. For a given spin and inclination angle, the shadow size and hence its average radius is larger than that of the Kerr black hole when $r_0> r_{0c}$ (see the discussion in Subsec. \ref{sec:shadows_RSV}). As a result, for a given $a$ and $\theta_o$, the fractional deviation parameter $\delta$ is larger than that of the Kerr black hole, i.e., $\delta>\delta_{Kerr}$ when $r_0>r_{0c}$. Therefore, with increasing $r_0$ beyond $r_{0c}$, the average shadow diameter and hence $\delta$ increases, and at some value $r_0=r_{0,max}$, $\delta$ becomes equal to the allowed upper bound $0.01$ (VLTI) or $0.05$ (Keck). Therefore, with a given $a$ and $\theta_o$, the shadow of the RSV metric is consistent with the observed shadow of Sgr A$^*$ for $r_0\leq r_{0,max}$.

\begin{figure}[h!]
\centering
\subfigure[\; $0\leq a/M \leq 0.8$]{\includegraphics[scale=0.42]{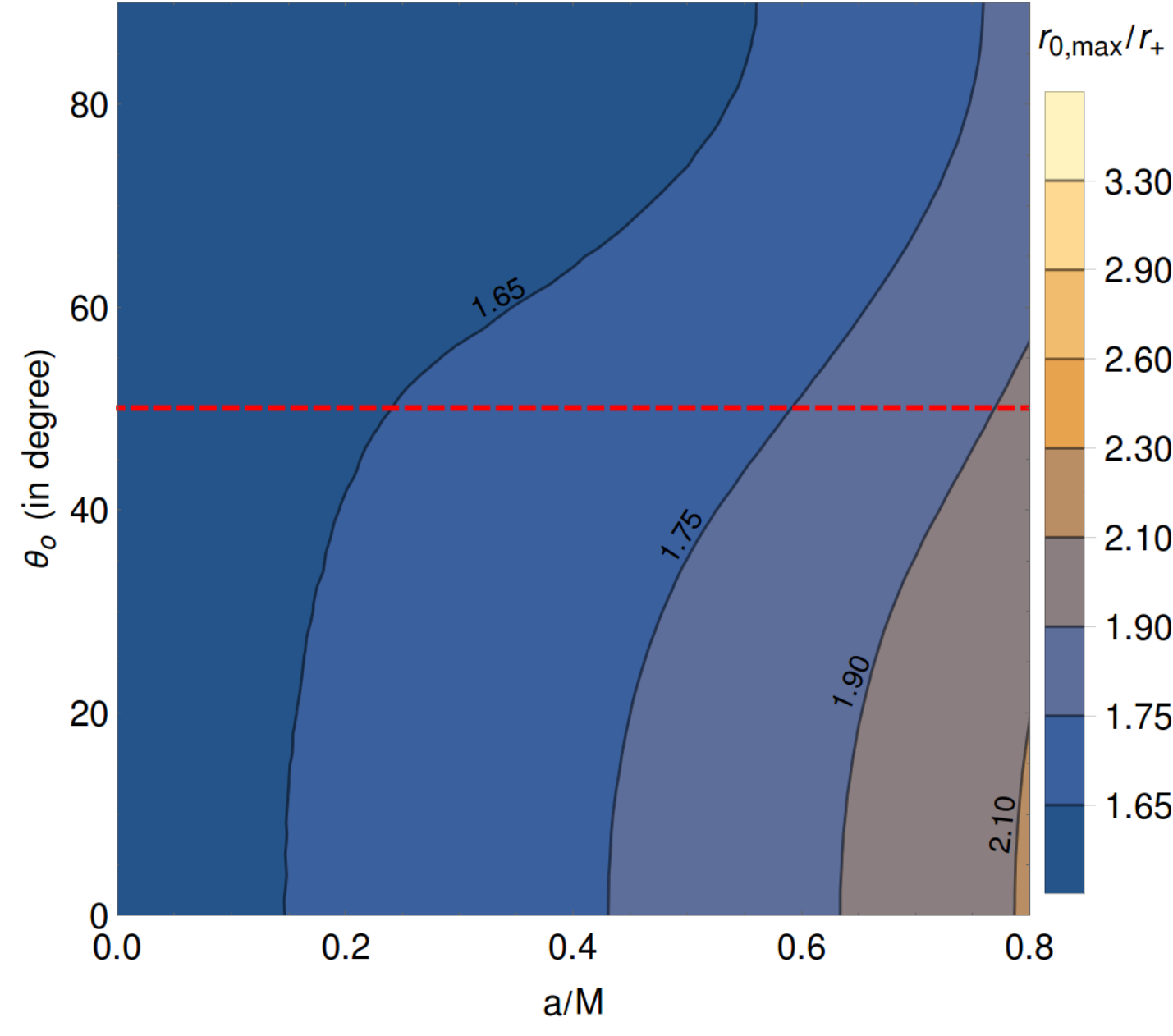}\label{fig:RSV_VLTI_delta1}}
\subfigure[\; $0.8\leq a/M \leq 1$]{\includegraphics[scale=0.40]{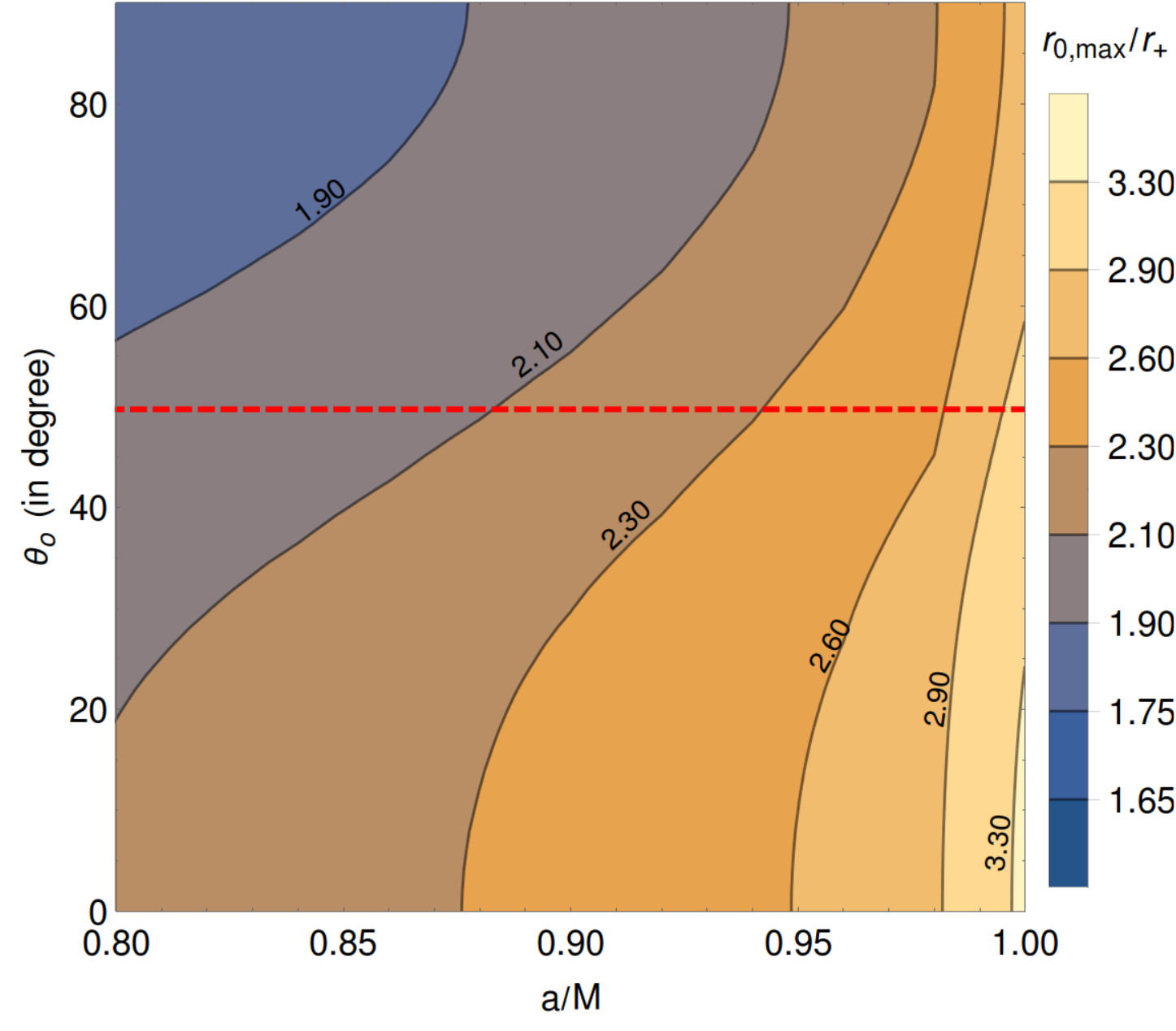}\label{fig:RSV_VLTI_delta2}}
\subfigure[\; $r_0=r_{0,max}$]{\includegraphics[scale=0.39]{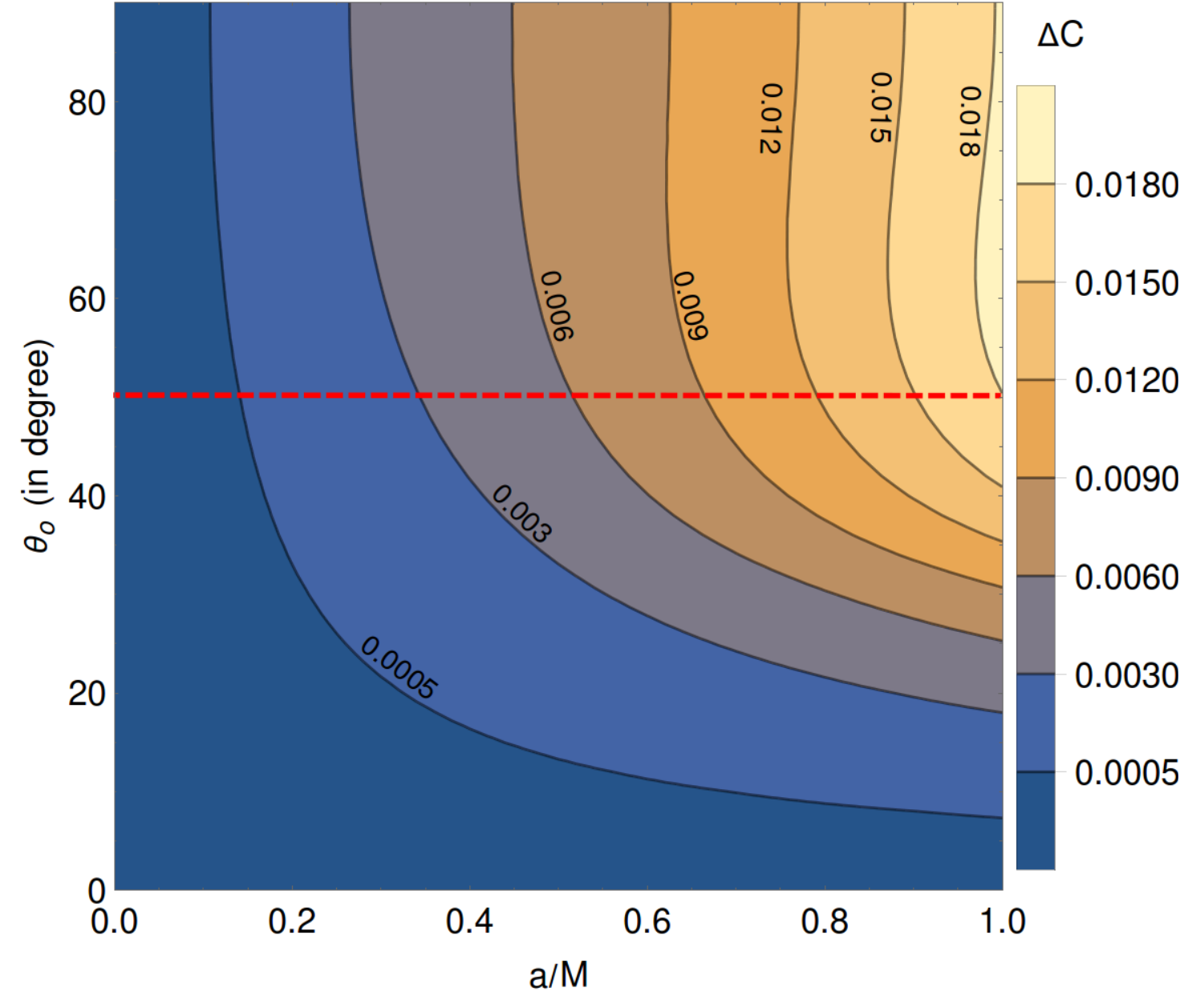}\label{fig:RSV_VLTI_deformation}}
\caption{Plots showing the maximum value $r_{0,max}$ of $r_0$ allowed for which the fractional deviation $\delta$ corresponds to the VLTI upper bound $0.01$ [(a) and (b)] and deviation from circularity $\Delta C$ at $r_0=r_{0,max}$ [(c)]. For a given spin and inclination angle, the shadow of the rotating SV metric satisfies the VLTI bound and hence, is consistent with the observed shadow of Sgr A$^*$ when $r_0\leq r_{0,max}$ (see discussion for details). Note that the maximum value of $\Delta C$ (at $r_0=r_{0,max}$) is less than that of the Kerr black hole shadow (compare with Fig. \ref{fig:Kerr_delta}). According to the EHT collaboration, an inclination angle higher than $50^\circ$ (i.e., $\theta_o>50^\circ$) is disfavoured. The red dashed curve shows $\theta_o=50^\circ$.}
\label{fig:RSV_VLTI_delta}
\end{figure}

\begin{figure}[h!]
\centering
\subfigure[\; $0\leq a/M \leq 0.8$]{\includegraphics[scale=0.42]{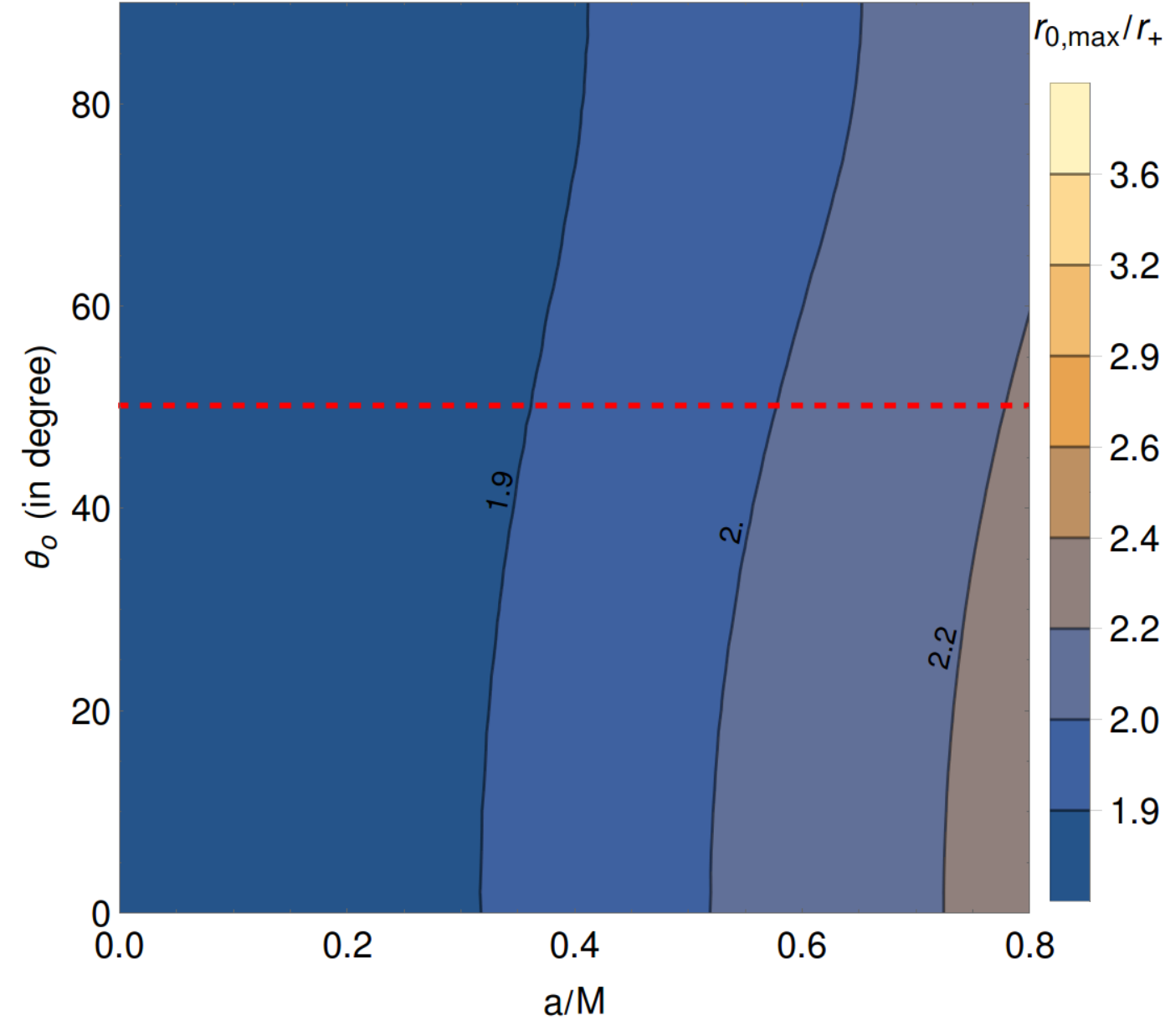}\label{fig:RSV_Keck_delta1}}
\subfigure[\; $0.8\leq a/M \leq 1$]{\includegraphics[scale=0.42]{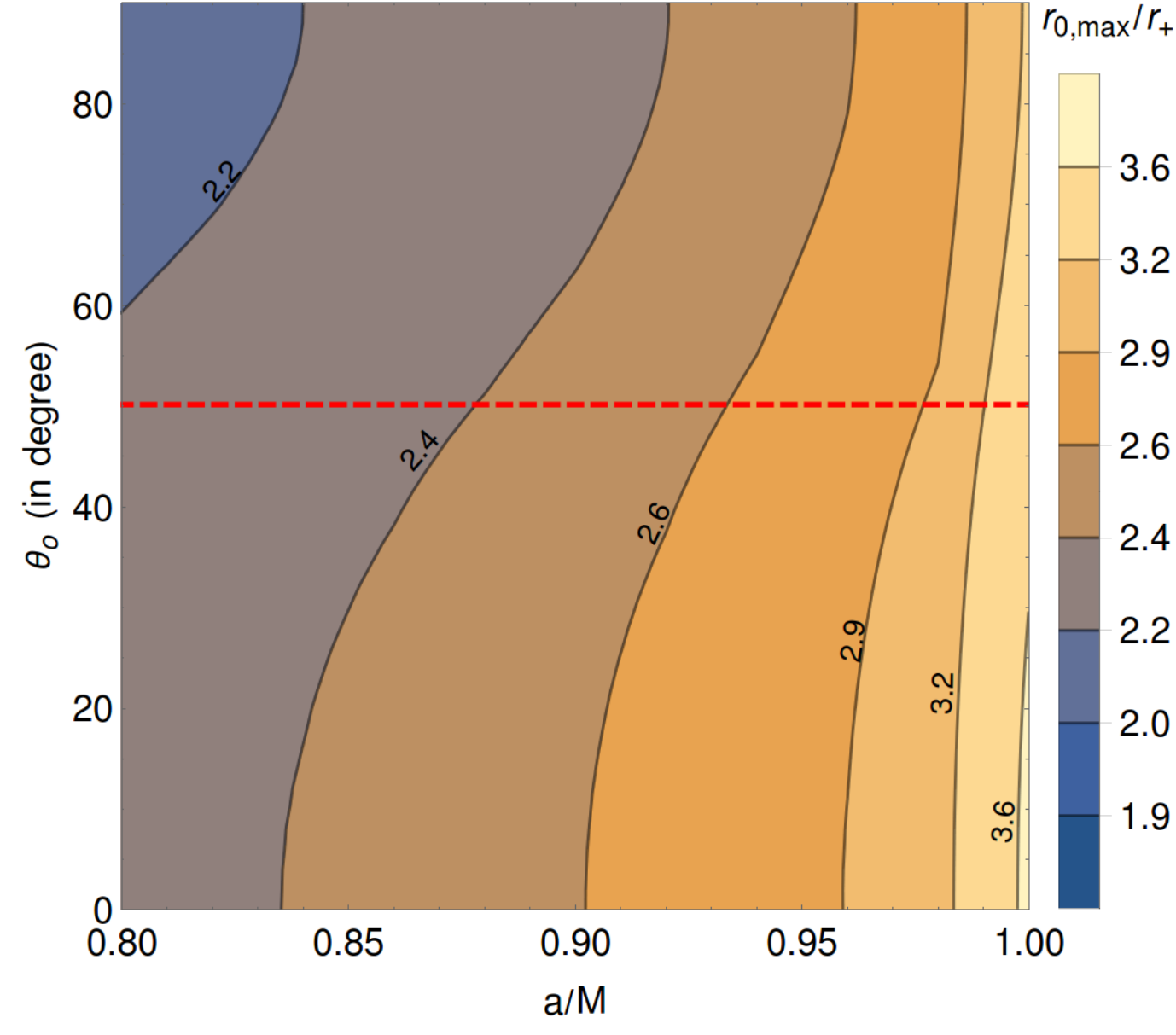}\label{fig:RSV_Keck_delta2}}
\subfigure[\; $r_0=r_{0,max}$]{\includegraphics[scale=0.39]{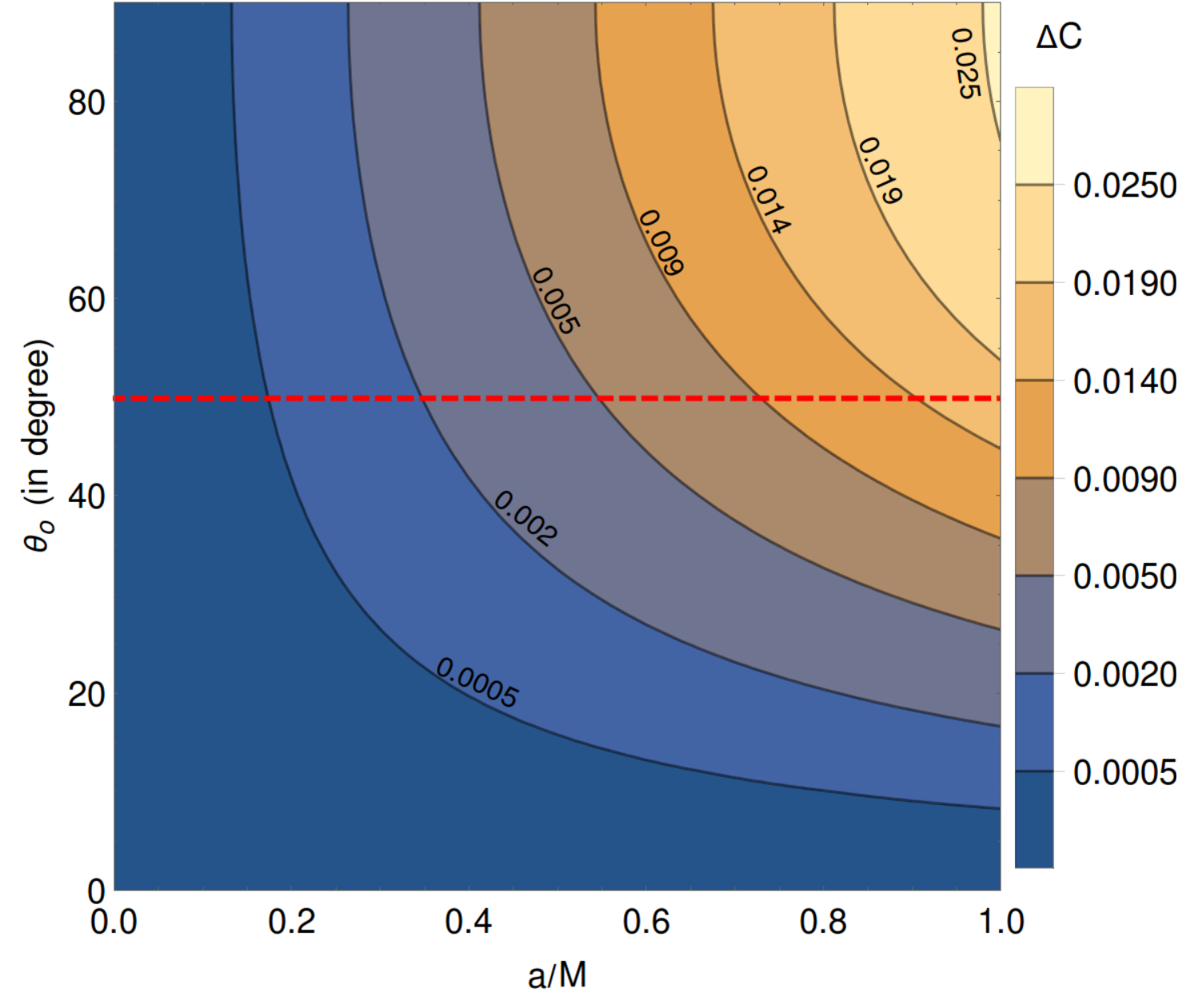}\label{fig:RSV_Keck_deformation}}
\caption{Plots showing the maximum value $r_{0,max}$ of $r_0$ allowed for which the fractional deviation $\delta$ corresponds to the Keck upper bound $0.05$ [(a) and (b)] and deviation from circularity $\Delta C$ at $r_0=r_{0,max}$ [(c)]. For a given spin and inclination angle, the shadow of the rotating SV metric satisfies the Keck bound and hence, is consistent with the observed shadow of Sgr A$^*$ when $r_0\leq r_{0,max}$ (see discussion for details). Note that the maximum value of $\Delta C$ (at $r_0=r_{0,max}$) is less than that of the Kerr black hole shadow (compare with Fig. \ref{fig:Kerr_delta}). According to the EHT collaboration, an inclination angle higher than $50^\circ$ (i.e., $\theta_o>50^\circ$) is disfavoured. The red dashed curve shows $\theta_o=50^\circ$.}
\label{fig:RSV_Keck_delta}
\end{figure}

Figures \ref{fig:RSV_VLTI_delta} and \ref{fig:RSV_Keck_delta} show $r_{0,max}$ as a function of the spin and inclination angle. When we consider the VLTI upper bound (Fig.~\ref{fig:RSV_VLTI_delta}), we find that, for $0\leq a/M\leq 1$, $1.636 r_{+}\lesssim r_{0,max}\lesssim 3.375 r_{+}$ when $\theta=0^\circ$ and $1.636 r_{+}\lesssim r_{0,max}\lesssim 3.006 r_{+}$ when $\theta=50^\circ$. However, when we consider the Keck upper bound (Fig.~\ref{fig:RSV_Keck_delta}), for $0\leq a/M\leq 1$, $1.848 r_{+}\lesssim r_{0,max}\lesssim 3.670 r_{+}$ when $\theta_o=0^\circ$ and $1.848 r_{+}\lesssim r_{0,max}\lesssim 3.458 r_{+}$ when $\theta_o=50^\circ$. Figures \ref{fig:RSV_VLTI_deformation} and \ref{fig:RSV_Keck_deformation} show the deformation $\Delta C$ from circularity at $r_0=r_{0,max}$. Note that maximum value of $\Delta C$ for the SV metric is less than that of the Kerr black hole shadow (compare with Fig. \ref{fig:Kerr_delta}). This is because the shadow for $r_0>r_{0c}$ in this case is more rounded than that of the Kerr black hole (see discussion in Subsec. \ref{sec:shadows_RSV}).

\subsection{$\gamma$-metric}

\begin{figure}[h!]
\centering
\subfigure[\; $0.5\leq \gamma\leq 1.5$]{\includegraphics[scale=0.43]{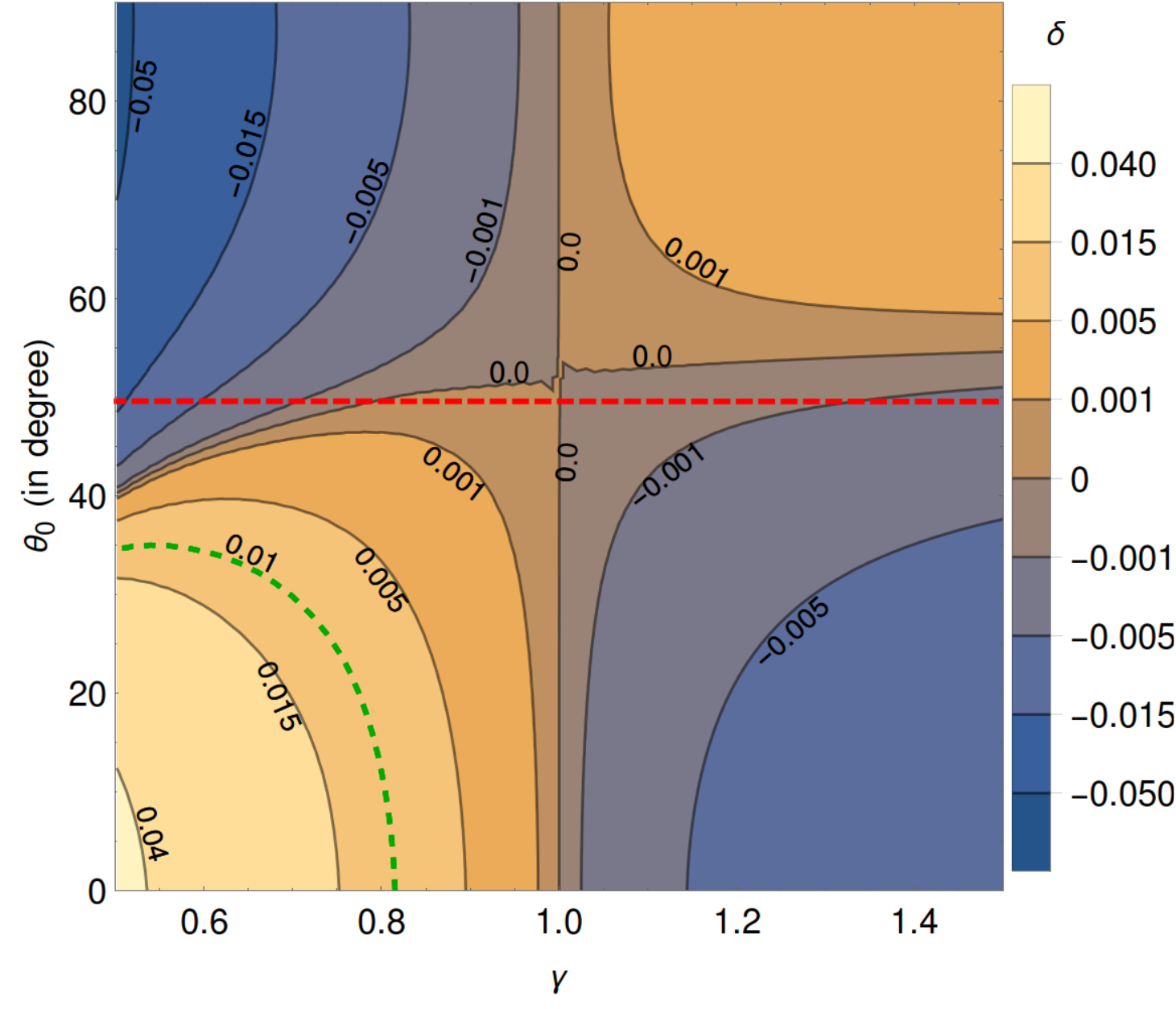}}
\subfigure[\; $1.5\leq \gamma\leq 8.0$]{\includegraphics[scale=0.42]{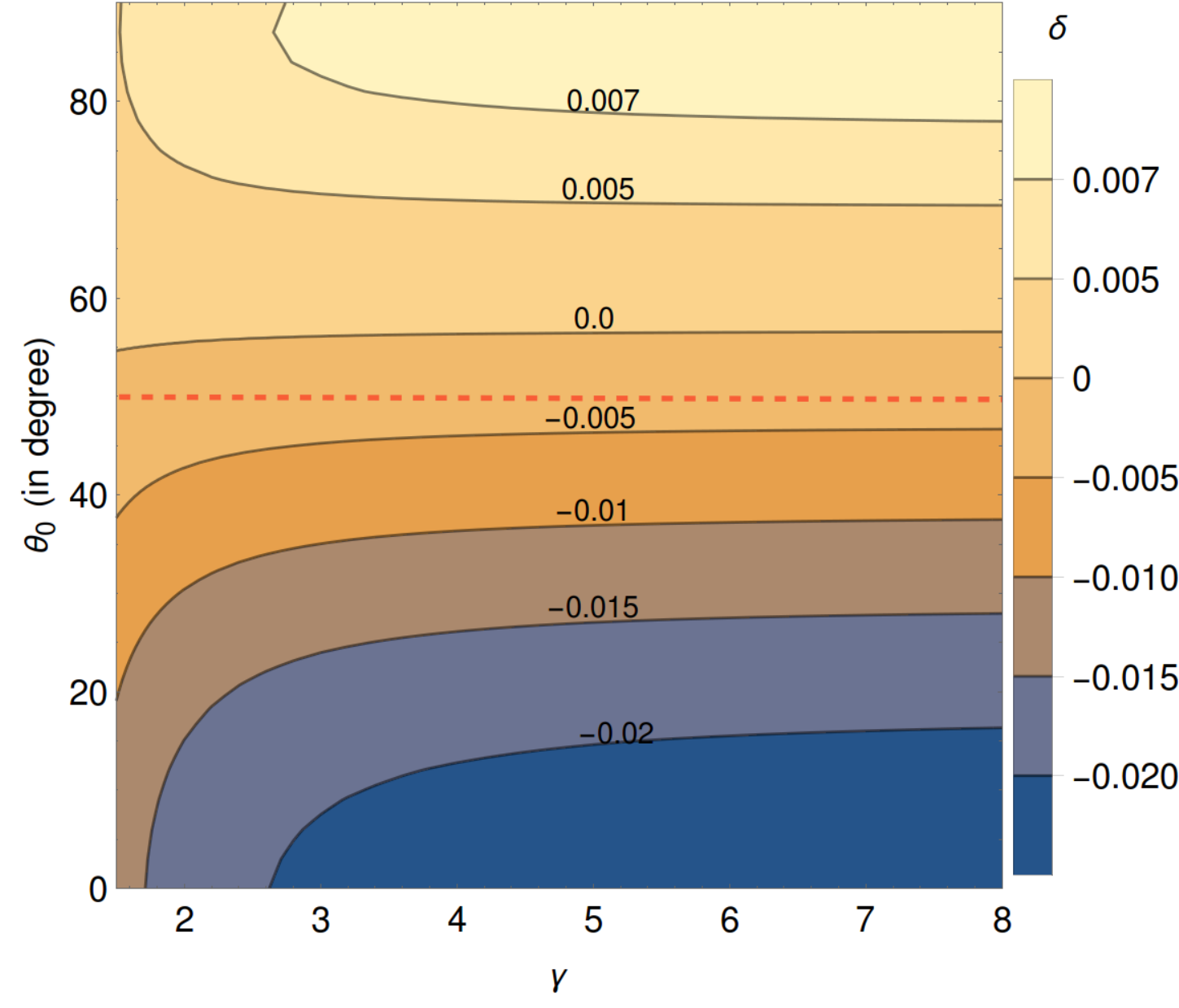}}
\subfigure[\; $\gamma\to \infty$ (GI)]{\includegraphics[scale=0.43]{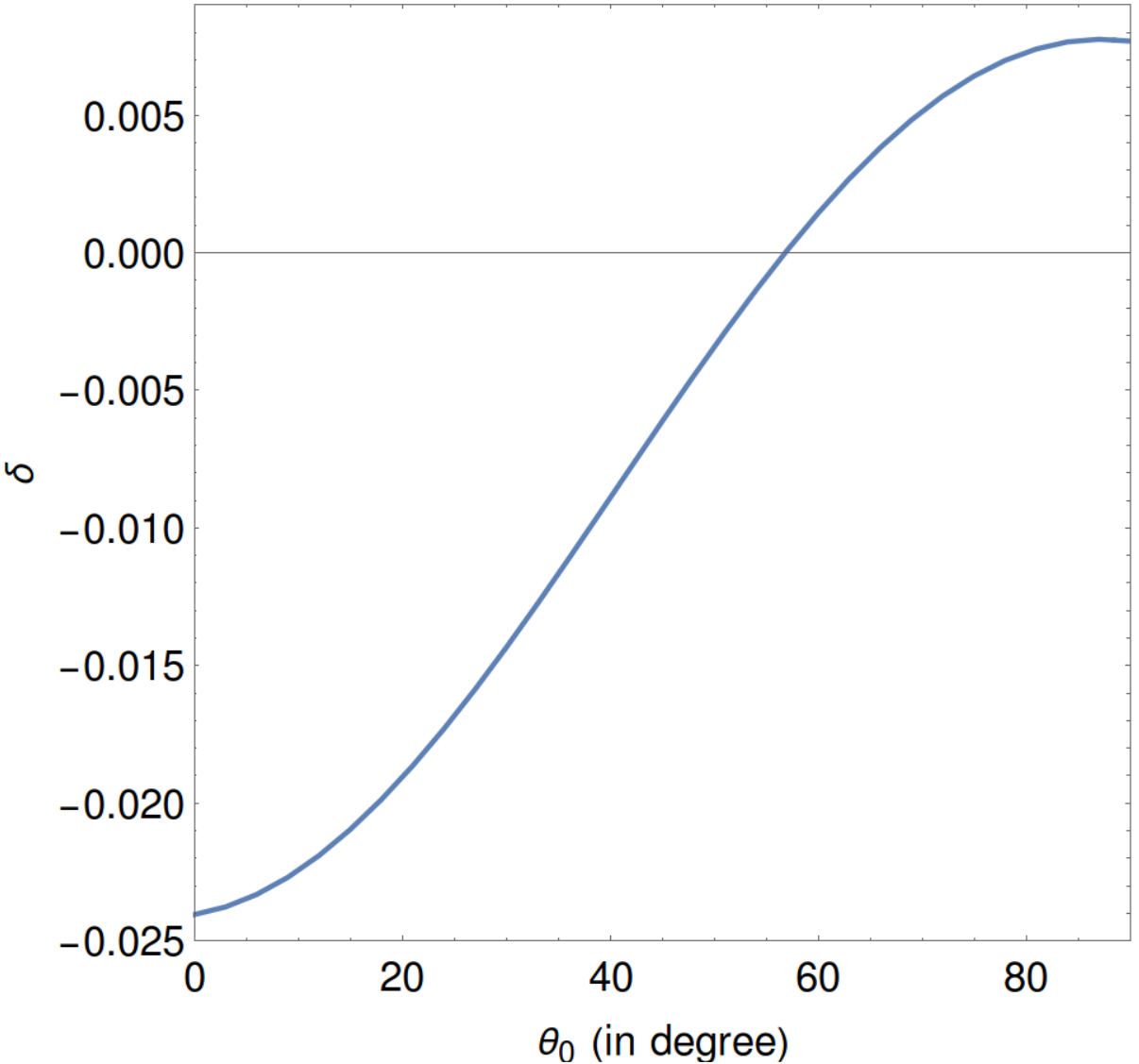}}
\caption{Plots showing the fractional deviation parameter $\delta$ for the shadow of the $\gamma$-metric [(a)-(c)]. The green dashed curve corresponds to the VLTI upper bound $0.01$. The parameter region right (left) to this curve is allowed (excluded) if we consider the VLTI bound. However, the Keck bound does not put any constraint on $\gamma$ as $\delta$ always remains within the bound (see discussion for details). The red dashed curve shows the inclination angle $\theta_o=50^\circ$.}
\label{fig:gamma_delta}
\end{figure}

Figure \ref{fig:gamma_delta} shows the fractional deviation parameter $\delta$ of the $\gamma$-metric shadow. Note that $\delta=0$ for $\gamma=1$ as expected. We have found that $\delta$ has a minimum value $\delta_{min}\simeq -0.063$ at $(\gamma,\theta_o)=(0.5,90^\circ)$ and a maximum value $\delta_{max}\simeq 0.047$ at $(\gamma,\theta_o)=(0.5,0^\circ)$. $\delta$ remains within these values, i.e., $-0.063\lesssim\delta\lesssim 0.047$ for any other values of $\gamma$ ($\geq 0.5$) and $\theta_o$. Therefore, the Keck bound does not put any constraint on $\gamma$, allowing all values $\gamma\geq 0.5$. However, the VLTI upper bound $0.01$ excludes some parameter region left to the green dashed curve in Fig. \ref{fig:gamma_delta}. The region right to this curve is allowed. Note that the green dashed curve touches the $\theta_o$-axis at $\theta_o\simeq 35^\circ$. Therefore, the VLTI bound puts a lower bound on $\gamma$ for inclination angle $\theta_o\lesssim 35^\circ$. For $\theta_o=0^\circ$, the allowed range is $\gamma\gtrsim 0.81$. With increasing $\theta_o$, the lower bound on $\gamma$ moves towards the minimum possible value $\gamma=0.5$ and coincides with $\gamma=0.5$ at $\theta_o\simeq 35^\circ$. For $\theta_o\gtrsim 35^\circ$, all possible values of $\gamma$ ($\geq 0.5$) are allowed. It is to be noted that $\gamma\geq 0.5$ is required for the $\gamma$-metric to cast a shadow.

\begin{figure}[]
\centering
\subfigure[\; $0.5\leq \gamma\leq 1.5$]{\includegraphics[scale=0.43]{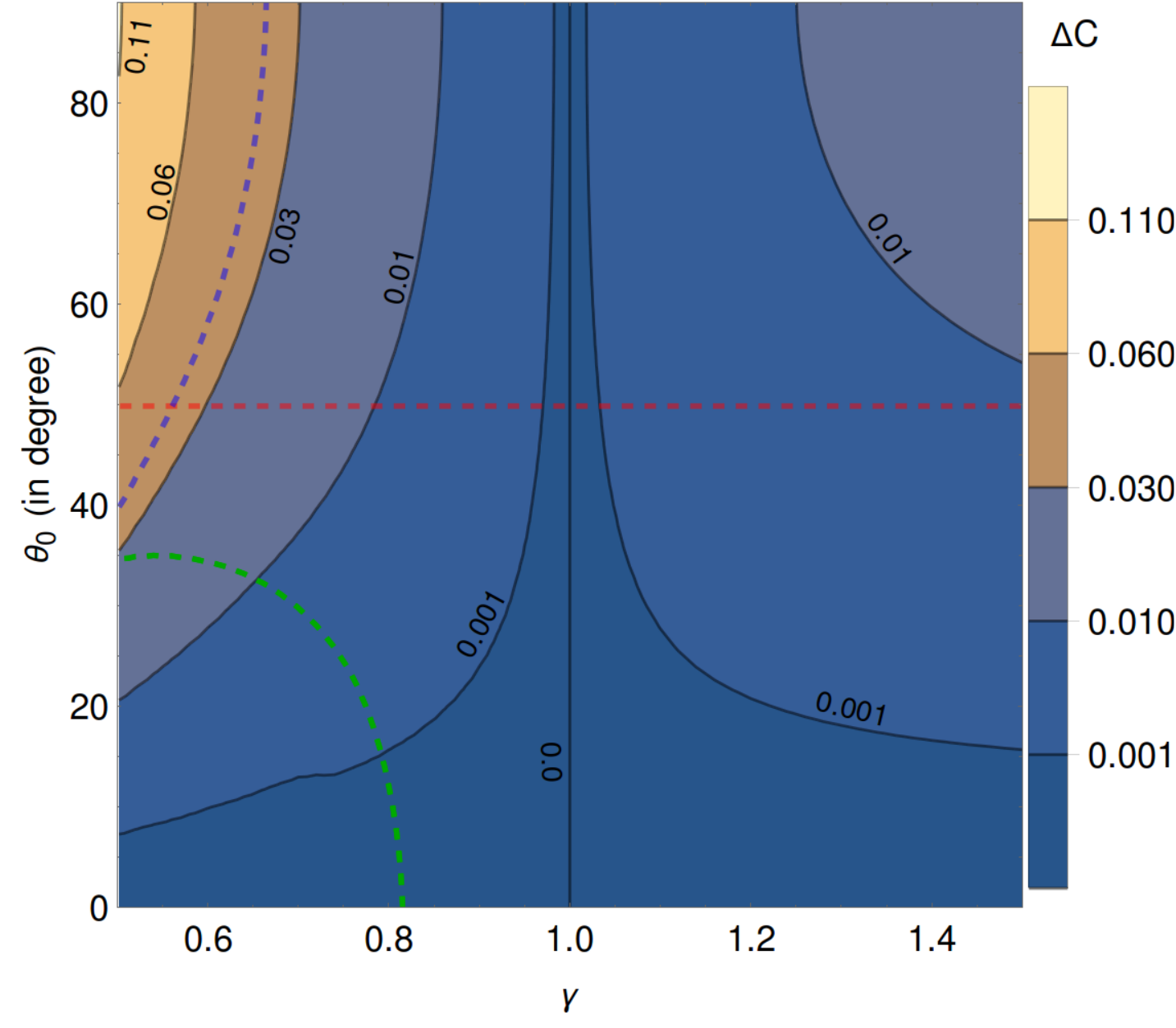}}
\subfigure[\; $1.5\leq \gamma\leq 8.0$]{\includegraphics[scale=0.43]{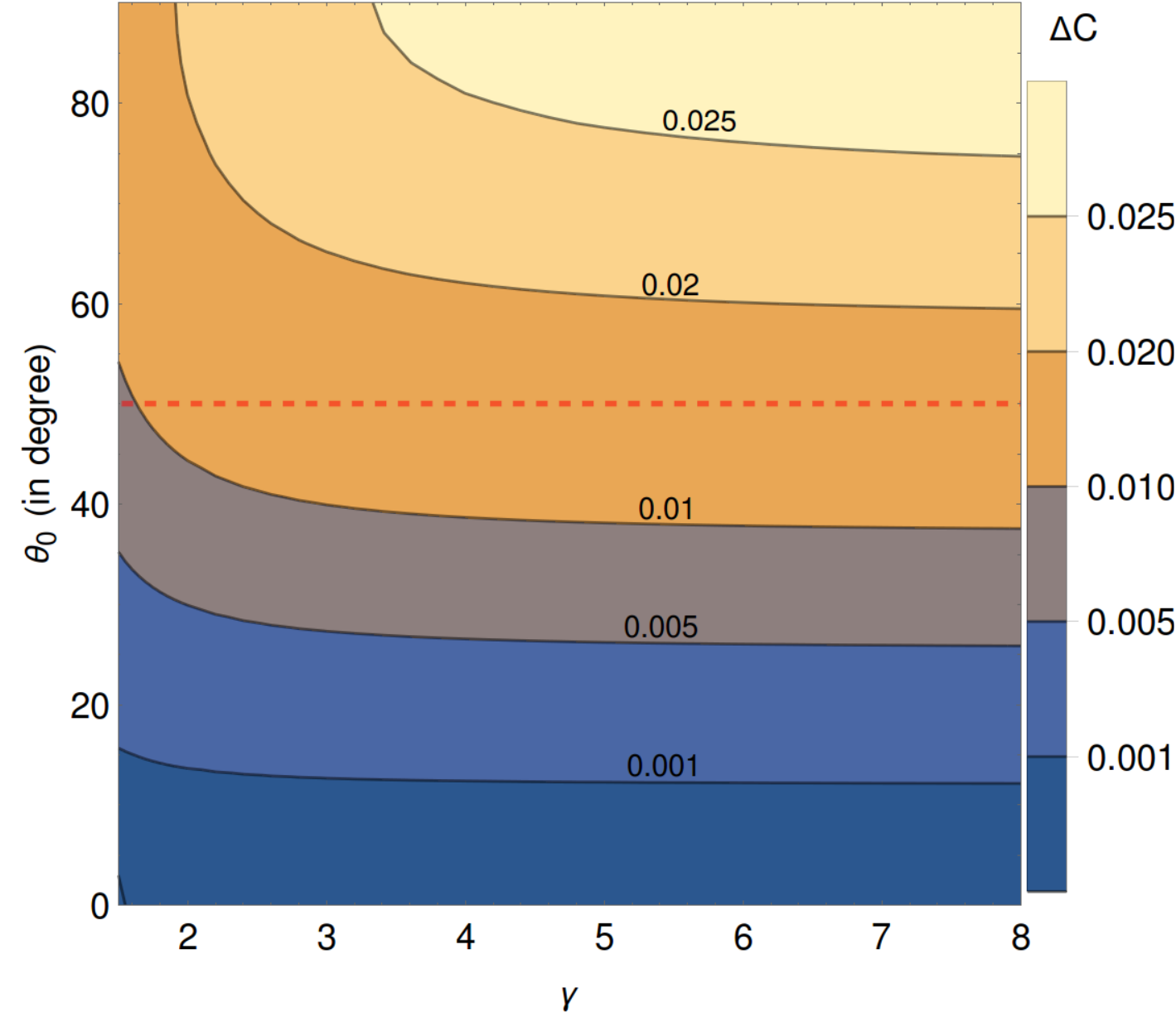}}
\subfigure[\; $\gamma\to \infty$ (GI)]{\includegraphics[scale=0.44]{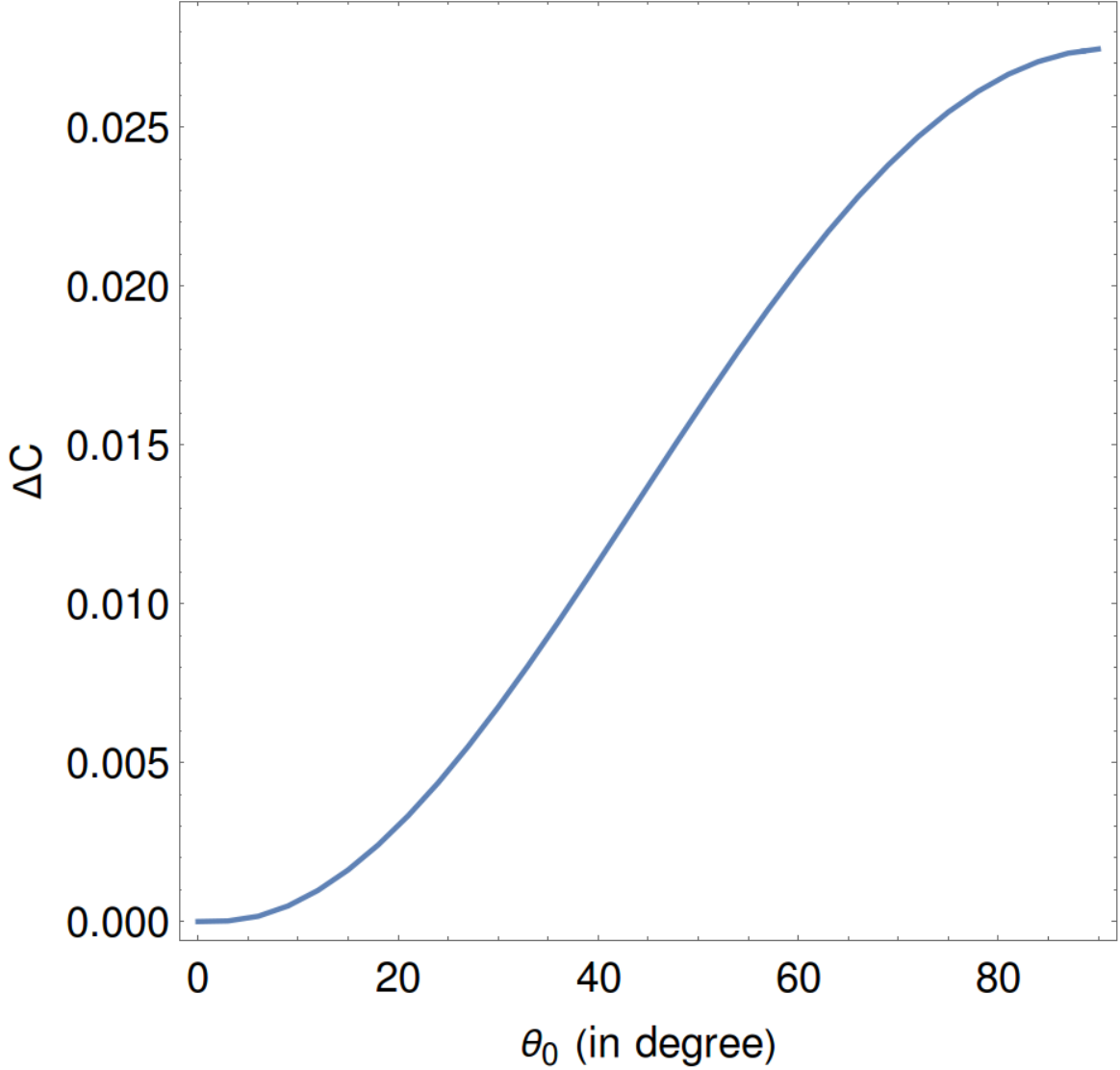}}
\caption{Plots showing the deviation $\Delta C$ from circularity of the shadow of the $\gamma$-metric [(a)-(c)]. The blue dashed curve corresponds to $\Delta C=0.0374$ which is the maximum deviation from circularity of the Kerr black hole shadow, considering all spins and $\theta_o\leq 50^\circ$. The green dashed curve corresponds to the VLTI upper bound $0.01$ for the fractional deviation parameter $\delta$ (see Fig. \ref{fig:gamma_delta}). The parameter region right (left) to this green curve is allowed (excluded) if we consider the VLTI bound for $\delta$. However, the Keck bound does not put any constraint on $\gamma$ (see discussion for details). The red dashed curve shows the inclination angle $\theta_o=50^\circ$.}
\label{fig:gamma_deformation}
\end{figure}

Figure \ref{fig:gamma_deformation} shows the deviation $\Delta C$ from circularity. Note that $\Delta C=0$ for $\gamma=1$ as expected. We have found that $\Delta C$ has a maximum value $\Delta C_{max}\simeq 0.116$ at $(\gamma,\theta_o)=(0.5,90^\circ)$ and remains below this value for all other values of $\gamma$ and $\theta_o$. Hence, the deviation from circularity can be upto $11.6\%$. However, if we restrict the inclination to $\theta_o\leq 50^\circ$, then the maximum deviation is $\Delta C_{max}\simeq 0.057$ which occurs at $(\gamma,\theta_o)=(0.5,50^\circ)$. The blue dashed curve in Fig. \ref{fig:gamma_deformation} marks the maximum $\Delta C$ that a Kerr black hole shadow can have when $\theta_o\leq 50^\circ$, considering all spins. Therefore, for $\theta_o\leq 50^\circ$, $\Delta C$ of the $\gamma$-metric shadow is always smaller than the maximum $\Delta C$ of the Kerr black hole shadow, except in the tiny region bounded by the red dashed curve, the blue dashed curve and the $\theta_o$ axis. Note that the EHT collaboration has not provided any constraint on $\Delta C$. Any constraint on $\Delta C$, together with that on $\delta$, might put more tighter constraint on $\gamma$.

\section{Conclusions}
\label{sec:conclusion}

The recent observations of the images and shadows of the supermassive compact objects Sgr A$^*$ and M87$^*$ at the hearts of Our Galaxy and the nearby galaxy M87, respectively, by the EHT collaboration have opened up a new window in observational astronomy to probe and test gravity and fundamental physics in the strong-field regime. It is commonly believed that the gravitational field around astrophysical compact objects is described by the Kerr geometry. The EHT data, therefore, can be used to test this hypothesis and put constraint on spacetime geometries which deviates from the Kerr geometry. 

In this paper, we have considered two black hole mimickers, namely the rotating Simpson-Visser (SV) black-bounce spacetime where a parameter $r_0$ measures deviation from the Kerr black hole geometry and the $\gamma$-metric where a parameter $\gamma$ measures deformation from the Schwarzschild black hole geometry. We have constrained these mimickers by comparing their shadows with the observed shadow of Sgr A$^*$. For the rotating SV metric, we have found out a upper bound $r_{0,max}$ on the parameter $r_0$ so that the fractional deviation parameter $\delta$ of its shadow satisfies the VLTI or Keck bound. For a given spin and inclination angle, the shadow of the rotating SV metric is consistent with the observed shadow of Sgr A$^*$ for $r_0\leq r_{0,max}$. For the $\gamma$-metric, the Keck bound on the fractional deviation parameter does not put any constraint on $\gamma$, allowing all possible values $\gamma\geq 0.5$. However, the VLTI bound puts a lower bound on $\gamma$ for inclination angle $\theta_o\lesssim 35^\circ$. For $\theta_o=0^\circ$, the allowed range is $\gamma\gtrsim 0.81$. With increasing $\theta_o$, the lower bound on $\gamma$ moves towards the minimum possible value $\gamma=0.5$ and coincides with $\gamma=0.5$ at $\theta_o\simeq 35^\circ$. For $\theta_o\gtrsim 35^\circ$, all possible values of $\gamma$ ($\geq 0.5$) are allowed.

It is to be noted that the EHT collaboration has not provided any bound on the deviation $\Delta C$ of the shadow from circularity or any tighter constraint on the spin and the inclination angle for Sgr A$^*$. These additional constraints, if available in future, together with the bound on the fractional deviation parameter $\delta$, might put more tighter constraints on the parameters $r_0$ and $\gamma$. It is to be noted also that we have considered the theoretically computed shadow size to be same as the observed central brightness depression or image shadow to constrain the mimickers. While the former depends only on spacetime geometry (through the unstable photon orbits), the latter additionally depends on the morphology of the emission region (plasma environment) surrounding the compact objects. Therefore, the latter may differ from the former. This has been discussed in \cite{Bronzwaer} by considering different GRMHD-based models of accreting black holes.

\section*{Acknowledgements}
This work is supported by the grant from the National Research Foundation funded by the Korean government, No. NRF-2020R1A2C1013266.

\end{document}